\documentclass[aps,prd,twocolumn,twoside,superscriptaddress,floatfix, nofootinbib]{revtex4-2}

\usepackage{amsmath}
\usepackage{mathtools}
\DeclareMathOperator{\arcsec}{arcsec}
\usepackage{aas_macros}
\usepackage{bm}
\usepackage{graphicx}
\usepackage{hyperref}
\usepackage{cleveref}
\usepackage[dvipsnames]{xcolor}
\usepackage{placeins}
\usepackage{float}
\usepackage{ulem}

\def\apjs{ApJS }               
\def\apjl{ApJ }                

\newcommand{\vl} { {\bm{\ell}} }

\newcommand{\beq} {\begin{equation}}
\newcommand{\eeq} {\end{equation}}
\newcommand{\bal} {\begin{aligned}}
\newcommand{\eal} {\end{aligned}}

\colorlet{violet}{black} 
\colorlet{red}{black}      

\begin{document}

\title{The Moving Lens Effect: Simulations, Forecasts and Foreground Mitigation}

\author{Ali Beheshti}
\email{ali.beheshti@pitt.edu}
\affiliation{Department of Physics and Astronomy, University of Pittsburgh, 3941 O’Hara Street, Pittsburgh, PA 15260, USA}
\affiliation{Pittsburgh Particle Physics, Astrophysics, and Cosmology Center (PITT PACC), University of Pittsburgh, Pittsburgh, PA 15260, USA}

\author{Emmanuel Schaan}
\email{eschaan@stanford.edu}
\affiliation{Kavli Institute for Particle Astrophysics and Cosmology,
382 Via Pueblo Mall Stanford, CA 94305-4060, USA}
\affiliation{SLAC National Accelerator Laboratory, 2575 Sand Hill Road Menlo Park, California 94025, USA}

\author{Arthur Kosowsky}
\email{kosowsky@pitt.edu}
\affiliation{Department of Physics and Astronomy, University of Pittsburgh, 3941 O’Hara Street, Pittsburgh, PA 15260, USA}
\affiliation{Pittsburgh Particle Physics, Astrophysics, and Cosmology Center (PITT PACC), University of Pittsburgh, Pittsburgh, PA 15260, USA}

\begin{abstract}

The peculiar motion of massive objects 
across the line of sight imprints a dipolar temperature anisotropy pattern on the cosmic microwave background known as the moving lens effect. 
This effect provides a unique probe of the transverse components of the peculiar velocity field, but has not yet been detected due to its small size. 
We implement and validate a stacking estimator for the moving lens signal using a galaxy catalog as a tracer of massive haloes combined with reconstructed velocities from the galaxy number density field.
Using simulations, we
forecast detection prospects for the moving lens signal from current and upcoming microwave background and galaxy surveys.
We demonstrate a new foreground mitigation  strategy \textcolor{violet}{likely} sufficient for current data sets,
and discuss various sources of systematic error and noise.
Upcoming galaxy surveys will provide high-significance statistical detections of the moving lens effect. 

\end{abstract}

\maketitle

\section{Introduction}

The cosmic microwave background (CMB) serves as a remarkable canvas that captures and preserves crucial information about the large-scale structure of the universe~\cite{2008RPPh...71f6902A}. Several distinct physical effects associated with galaxy cluster haloes have been detected:
gravitational lensing~\cite{2006PhR...429....1L} (depending on the total halo mass distribution); the  
late-time Integrated Sachs-Wolfe (ISW) effect~\cite{1967ApJ...147...73S} and the Rees-Sciama effect \cite{1968Natur.217..511R} (due to the time evolution of cluster potentials); 
the thermal Sunyaev-Zel'dovich (tSZ) effect
\cite{1972CoASP...4..173S}
(proportional to ionized gas pressure along the line of sight); and the kinematic Sunyaev-Zel'dovich (kSZ) effect \cite{1980ARA&A..18..537S,1980MNRAS.190..413S} (depending on the ionized gas momentum along the line of sight).

The last of these has been exploited in recent years to detect the motions of objects at cosmological distances along the line of sight.
The kSZ effect produces a nearly blackbody distortion of the cosmic microwave background in the direction of a cluster proportional to the line-of-sight velocity, analogous to the Doppler effect. Its small amplitude (typically a few micro-Kelvin for a galaxy cluster with mass of $5\times 10^{14}\,M_\odot$) makes a detection for individual clusters challenging \cite{2013ApJ...778...52S}. 
Statistical kSZ detections have been accomplished with two distinct techniques: mean pairwise velocities \cite{PhysRevLett.109.041101,Bernardis_2017,PhysRevD.104.043502}, which detects the slight tendency for random pairs to be moving towards each other, and by stacking the sky signal around all clusters, with the sign of each contribution determined using a velocity template estimated from the galaxy density field \cite{PhysRevD.103.063513}. 
For a given galaxy catalog and microwave background map, the signal-to-noise ratio (SNR) of both techniques is similar in principle \cite{2018arXiv181013423S}.

The microwave background also provides a unique avenue for measuring velocities {\it transverse} to the line of sight, known as the moving lens effect \cite{1983Natur.302..315B,1986Natur.324..349G,1993ApJ...415..459P,1998A&A...334..409A}. A small dipole distortion of the microwave background temperature in the direction opposite the transverse component of the peculiar velocity results from the interplay between the transverse motion and the gradient of the gravitational potential. Remarkably, the latter quantity corresponds precisely to the lensing deflection, an observable that can be measured independently. 
Consequently, combining lensing and moving lens measurements provides a path to obtaining the transverse velocity component. 
In contrast, the kSZ effect is proportional to the optical depth to Thomson scattering through the halo, 
which is challenging to measure to high precision independently, without additional astrophysical assumptions.

While the moving lens effect has not yet been detected, 
current and forthcoming CMB surveys such as the Atacama Cosmology Telescope (ACT  \cite{2007ApOpt..46.3444F,2011ApJS..194...41S,2016ApJS..227...21T,2016JLTP..184..772H}), 
Simons Observatory (SO \cite{2019JCAP...02..056A}), 
and the Stage-IV CMB experiment (S4 \cite{2016arXiv161002743A}), 
combined with ever-expanding galaxy catalogs from large scale structure \textcolor{red}{(LSS)} surveys such as Baryon Oscillation Spectroscopic Survey (BOSS \cite{2014ApJS..211...17A,2017MNRAS.464.1640S,2017MNRAS.470.2617A}), 
Dark Energy Spectroscopic Instrument (DESI \cite{2016arXiv161100036D}) 
and the Vera C.~Rubin Observatory Legacy Survey of Space and Time (LSST \cite{2009arXiv0912.0201L}), provide data sets of sufficient size that statistical detection becomes possible. 
Moreover, the accuracy and precision of photometric redshifts expected from forthcoming surveys such as Rubin Observatory \cite{2018arXiv180901669T,2022ARA&A..60..363N}  
will enable the use of far larger galaxy catalogs. Notably, the estimation of velocity components across the line of sight is less susceptible to photo-z errors than the component along the line of sight \cite{2023arXiv231212435R}.

This study revisits predictions related to the moving lens effect \cite{2019ApJ...873L..23Y,2004A&A...419..439R,2007A&A...467..411M,2019PhRvL.123f1301H,2021PhRvD.103d3536H,2021PhRvD.104h3529H,HotPaol}, using well-known methodologies employed previously in stacked kSZ measurements with velocity reconstruction \cite{2021PhRvD.103f3513S}. 
We underscore several salient details of the moving lens signal, including its relatively large angular extent of the signal and the importance of the two-halo term.
We update and correct the public code \texttt{AstroPaint}
~\cite{2020JOSS....5.2608Y} which
generates mock moving lens maps on the curved sky\footnote{The updated \texttt{AstroPaint} version is available at \url{https://github.com/ali-beheshti/Astro-Paint}}
, and introduce a new public code for faster generation of mock moving lens maps on the flat sky\footnote{\url{https://github.com/EmmanuelSchaan/ThumbStack/tree/moving_lens}\label{TS}}.
Careful consideration of convergence properties of the signal is paramount to avoiding significant biases in the analysis. 
Similarly to Ref.~\cite{HotPaol}, we implement an estimator which performs oriented stacks of the microwave background intensity along the direction of transverse velocities of galaxy clusters to estimate the average angular profile of the moving lens temperature anisotropy. We use transverse components of the reconstructed velocities from the galaxy number density field, similar to the kSZ measurements. This stacking estimator is integrated into the public 
\texttt{ThumbStack}\footnotemark[\getrefnumber{TS}] code. 

Utilizing this framework, we provide realistic 
projections for forthcoming experiments. 
Identifying and confirming extragalactic foreground biases, as highlighted in Ref.~\cite{2023PhRvD.108h3508H}, leads us to advocate for a simple deprojection technique using the galaxy catalog. 
We verify the efficacy of this method in mitigating such biases.
Furthermore, we argue that CMB lensing, responsible for local dipoles in CMB maps 
aligned with the local microwave temperature gradient, does not bias the moving lens estimator. 
Finally, we consider how miscentering in the stacking analysis can degrade the signal to noise ratio of the moving lens estimates. 
 
This paper is structured as follows. In Section~\ref{mlReview}, we review the moving lens effect and provide intuition for the point mass and dark matter halo cases. Section~\ref{simMaps} describes simulation of the moving lens signal maps. Section~\ref{pipeline} gives details on the properties of our aperture filter and dipole estimator, and presents model assumptions about cosmic microwave background experiments and large scale structure surveys.
Details on noise simulations, plus detection forecasts and a discussion about mitigating biases, are given in Sec.~\ref{results}. Finally,  Section~\ref{conclusion} looks ahead to a first detection of the moving lens effect and its potential cosmological impacts.

\section{General Derivation of the Moving Lens Effect} \label{mlReview}

Consider the path of a photon through the gravitational potential $\Phi$ of a moving galaxy cluster. 
Since the gravitational potential is time-dependent, the photon energy varies, such that
\begin{equation}
    \frac{\delta T(\bm{n})}{T}= \frac{2}{c^2} \int_{\eta_{*}}^{\eta_{0}} \partial_\eta \Phi(c(\eta - \eta_0)\bm{n},\eta) \; d\eta, 
\end{equation}
where $\eta_0$ and $\eta_*$ are the conformal time today and at the last scattering surface, respectively. 
The unit vector $\bm{n}$ is along the line of sight. 
In the case of the moving lens effect, the time dependence of the cluster gravitational potential is due to the motion of its center, so 
$\partial_\eta \Phi = - \bm{\nabla} \Phi \cdot \bm{v}$, 
where the gradient is taken with respect to comoving coordinates and the peculiar velocity $\bm{v}=d\bm{x}/d\eta$ is derivative of the halo's comoving coordinates with respect to conformal time. 
Then
\begin{equation} 
    \frac{\delta T(\bm{n})}{T} = - \frac{2}{c^2} \int_{\chi_0}^{\chi_*} \bm{\nabla} \Phi(\bm{n}) \cdot \frac{\bm{v}(\bm{n})}{c} d\chi \equiv \bm{\beta}(\bm{n}) \cdot \frac{\bm{v}(\bm{n})}{c}, 
\label{eq:lensing_deflection}
\end{equation}
where $\chi$ is the comoving distance from us to the photon on its trajectory. We recognize
$\bm\beta\equiv-\frac{2}{c^2}\int d\chi \; \bm{\nabla}\Phi$ 
is the deflection of the photon path due to gravitational lensing by the cluster, pointing towards the center of cluster. 
Hence the moving lens signal is a temperature dipole, oriented along the direction of the peculiar velocity across the line of sight.
This is similar to the lensing dipole signal,
$\delta T / T =\left( D_{\textcolor{red}{\mathrm{ls}}}/D_{\textcolor{red}{\mathrm{s}}} \right){\bm\beta}\cdot {\bm\nabla} T$, 
where $D_{\textcolor{red}{\mathrm{ls}}}$ and $D_{\textcolor{red}{\mathrm{s}}}$ are the comoving distances between the lens and source, and between the source and observer.
The difference between the moving lens dipole and the usual lensing dipole are only their amplitudes and orientations.
The lensing dipole is oriented along the unlensed temperature gradient, whereas the moving lens dipole is oriented along the cluster motion.
In particular, the radial profiles of both dipole signals are the same, determined by the deflection amplitude profile $|\bm{\beta}|$, i.e. the profile of the gradient of the gravitational potential. The microwave polarization map also traces the cluster lensing signal while being insensitive to transverse velocity; sufficiently good polarization maps can increase the detection efficiency of the transverse lensing signal by separating the lensing and transverse velocity components of the temperature dipole.

\subsection{Spherically Symmetric Cluster}

For a spherically symmetric cluster, the time dependence of the potential is given by
\begin{equation}
    \Phi(\bm{r},\eta)=\Phi(|\bm{r}-\bm{r_0}(\eta)|),
\end{equation}
where $\bm{r_0}(\eta)$ and $\bm{r}$ are the comoving positions of the cluster center and the photon, respectively. Hence:
\begin{align}
    \partial_\eta \Phi(\bm{r},\eta) =\;& \frac{d\Phi(|\bm{r}-\bm{r}_0(\eta)|)}{dr} \frac{\bm{r}-\bm{r}_0(\eta)}{|\bm{r}-\bm{r}_0(\eta)|} \cdot \frac{d(\bm{r}-\bm{r}_0(\eta))}{d\eta} \nonumber \\ 
    \equiv\;& - \Phi'(r) \frac{\bm{r}}{r} \cdot \bm{v}_0,
\end{align}
where $\bm{v_0}= d\bm{r}_0/d\eta$ is the peculiar velocity. 
In the second equality, we 
redefine $\bm{r} \equiv \bm{r}- \bm{r}_0$.
Vector $\bm{r}$ connects the halo center to the photon.
Therefore, the energy variation of the photon reduces to:
\begin{equation}
     \frac{\delta T}{T} = - \frac{2}{c^2} \int_{\chi_0}^{\chi_*} d\chi \; \Phi'(r) \frac{\bm{r}}{r}\cdot \frac{\bm{v}_0}{c} 
\end{equation}
In this integral, the component of $\bm{v_0}$ along the line of sight leads to a vanishing contribution by symmetry. 
The moving lens effect is therefore only sensitive to $\bm{v}_\perp$, the velocity perpendicular to the line of sight. 
Converting from comoving distance $\chi$ to comoving spherical radius $r$ gives
\begin{equation}\label{signal}
     \frac{\delta T}{T} = -\frac{4}{c^2} \left(\frac{\bm{v}_0}{c} \cdot \bm{r}_\perp\right)  \int_{r_\perp}^\infty dr \; \frac{\Phi'(r)}{\sqrt{r^2 - r^2_\perp}},
\end{equation}
where $r_\perp$ is the transverse comoving distance from the halo center to the trajectory of the photon.

\subsection{Point mass features}

Studying the point source case is helpful to understand the radial dependence of the signal far away from the massive object.
For a point source, the matter density is 
\beq
\rho (\bm{x}) = m\ \delta^D (\bm{x}),
\eeq
where $\delta^D (\bm{x})$ is the Dirac delta.
Such that the matter overdensity field is simply
\beq
\delta_m = \frac{m}{\bar{\rho}} \delta^D - 1,
\eeq
which can be plugged into the Poisson equation to give the gravitational potential $\Phi$. 
The Poisson equation in terms of density per unit comoving volume $\rho_m$ and Laplacian with respect to comoving coordinates reads
\begin{equation}
    \nabla^2 \Phi = 4 \pi G  \frac{\overline{\rho}_{m(z=0)} \delta_m}{a},
\end{equation}
where $a$ is the scale factor, $\overline{\rho}_{m(z=0)}$ is the mean matter density per unit comoving volume.
For the point source the Poisson equation yields the gravitational potential in terms of comoving radius as
\begin{equation}
    \Phi(r)= -\frac{4 G m}{a r}.
\end{equation}
Plugging its derivative $\Phi'(r)= Gm/a r^2$ into Eq.~(\ref{signal}) leads to the temperature distortion
\begin{equation} \label{pointmass}
\begin{split}
        \frac{\delta T}{T} &= -\frac{4Gm}{c^2 a r} \frac{v_0}{c} \cos{\phi} \\
        &= \beta(r) \frac{v_0}{c} \cos{\phi},
\end{split}
\end{equation}
where $\beta(r)= -4GM/c^2 ar = -2R_{\textcolor{red}{\mathrm{Sch}}}/ar$ is the familiar general relativity result for deflection angle of a photon traveling near a point mass \cite{2019sgai.book.....C} \textcolor{red}{and $R_{\mathrm{Sch}} = 2GM/c^2$ is the Schwarzschild radius.}
The signal thus scales as $1/r$, just like the lensing deflection.
Interestingly, this scaling implies that the integrated effect in a circular aperture, once the angular dependence is accounted for, keeps growing indefinitely as the aperture size increases.
This contrasts with, for example, the Sunyaev-Zel'dovich effects, where most of the integrated signal is concentrated within a few virial radii of the object.
We therefore expect the optimal filter or aperture for the moving lens effect to be wider, cut off only by the increasing noise in larger apertures.

\subsection{NFW mass profile}

In the case of an NFW density profile characterized by virial mass $M_{\text{vir}}$ and virial radius $R_{\text{vir}}$\cite{1997ApJ...490..493N}, the density profile is given by
\begin{equation}
    \rho(x)\equiv\frac{\rho_s}{x(1+x)^2}, \; \; \; x \equiv \frac{r}{r_s}
\end{equation}
where $r_s =R_{\text{vir}}/c$ is the scale radius, $c$ is the concentration parameter, $\rho_s \equiv M_{\text{vir}}/[{4\pi r_s^3(\ln(1+c)-c/(1+c))]}$ is the characteristic density,  and all quantities are comoving.
Using the Poisson equation we obtain the gravitational potential of the cluster as 
\begin{equation}
    \Phi(x)= -4 \pi G \frac{\rho_s r_s^2}{a}  \; \frac{\ln(1+x)}{x},
\end{equation}
and hence its derivative with respect to comoving radius:
\begin{align}
    &\Phi'(r) = \frac{d\Phi(r)}{dr}= \frac{1}{r_s}\frac{d\Phi(x)}{dx} = 4 \pi G \frac{\rho_s r_s}{a} \; u(x), \nonumber \\ 
    &u(x) = \frac{\ln(1+x)}{x^2} - \frac{1}{x(1+x)}.
\end{align}
Thus, the corresponding signal caused by the moving cluster on the sky can be obtained from Eq.~(\ref{signal}) as
\begin{align}\label{signal_final}
    &\frac{\delta T}{T}(x) = -\frac{16 \pi G}{c^2 } \frac{\rho_s r_s^2}{a} \frac{\bm{v}_\perp}{c} \cos{(\phi)} f(x), \nonumber \\
    &f(x) = \frac{1}{x} \left[ \left|\frac{\arcsec(x)}{\sqrt{x^2-1}}\right| + \ln(x/2) \right].
\end{align}

\begin{figure}[h]
    \centering    \includegraphics[width=.4\textwidth]{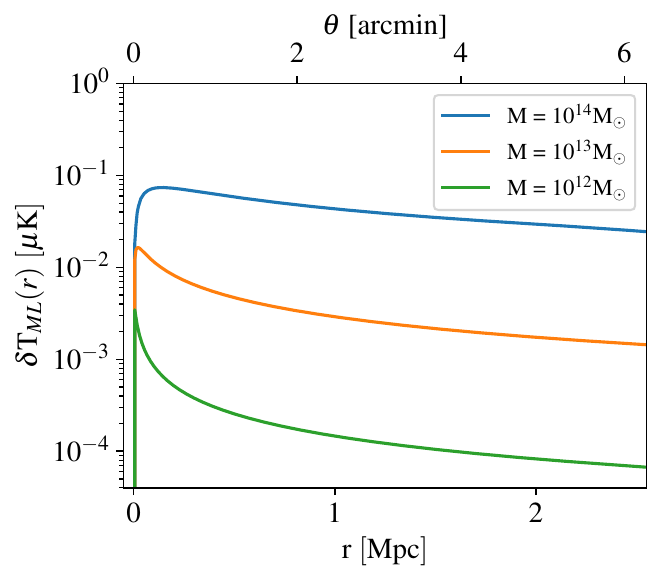}\\
    \caption{Amplitude of the temperature anisotropy caused by the moving lens signal in the direction of clusters velocity for three clusters with various masses, with $v_{\perp} = 200$ km/s and $z = 0.57$.
    This is similar to, but lower than the tSZ/kSZ signals of the same halos, and much smaller than the RMS primary CMB fluctuations of 110~$\mu$K.
    }
    \label{fig:signal3clusters}
\end{figure}

To give intuition for the size of this signal, Fig.~\ref{fig:signal3clusters} show the amplitude of the moving lens effect caused by three moving clusters with various masses, $v_\perp = 200$km/s and $z = 0.57$.
The NFW mass density profile, integrated over all space, has infinite mass.
In Appendix \ref{app:nfw_truncation_radius}, we display the large effect of the choice of truncation radius on the predicted moving lens signal. 
For the purpose of this study, we truncate the NFW density profile at the virial radius of each halo, such that the mass enclosed is exactly the virial mass.
At distances much greater than the virial radius (which is the finite extent of our mass profiles), the signal scales inversely with radius $r$, as with the point mass.

\section{Simulated signal maps}\label{simMaps}

The analytical derivations in the previous section are sufficient to simulate maps of the moving lens signal for any catalog of galaxies. 
\textcolor{red}{We introduce a new code for the efficient generation of moving lens mock maps on a flat sky patch, which is publicly available as part of the \texttt{ThumbStack} library. Additionally, we have updated the public \texttt{AstroPaint} code with corrections relevant to this application, including: (1) correcting typos in the NFW halo concentration parameters, (2) updating the expression for $\rho_\text{crit}$ to account more accurately for its redshift evolution, and (3) modifying the NFW halo deflection angle and moving lens temperature anisotropy expressions to simulate the effects of a truncated NFW matter density at a specified radius.} 

\begin{figure*}[htbp]
    \centering
    \includegraphics[width=1\textwidth]{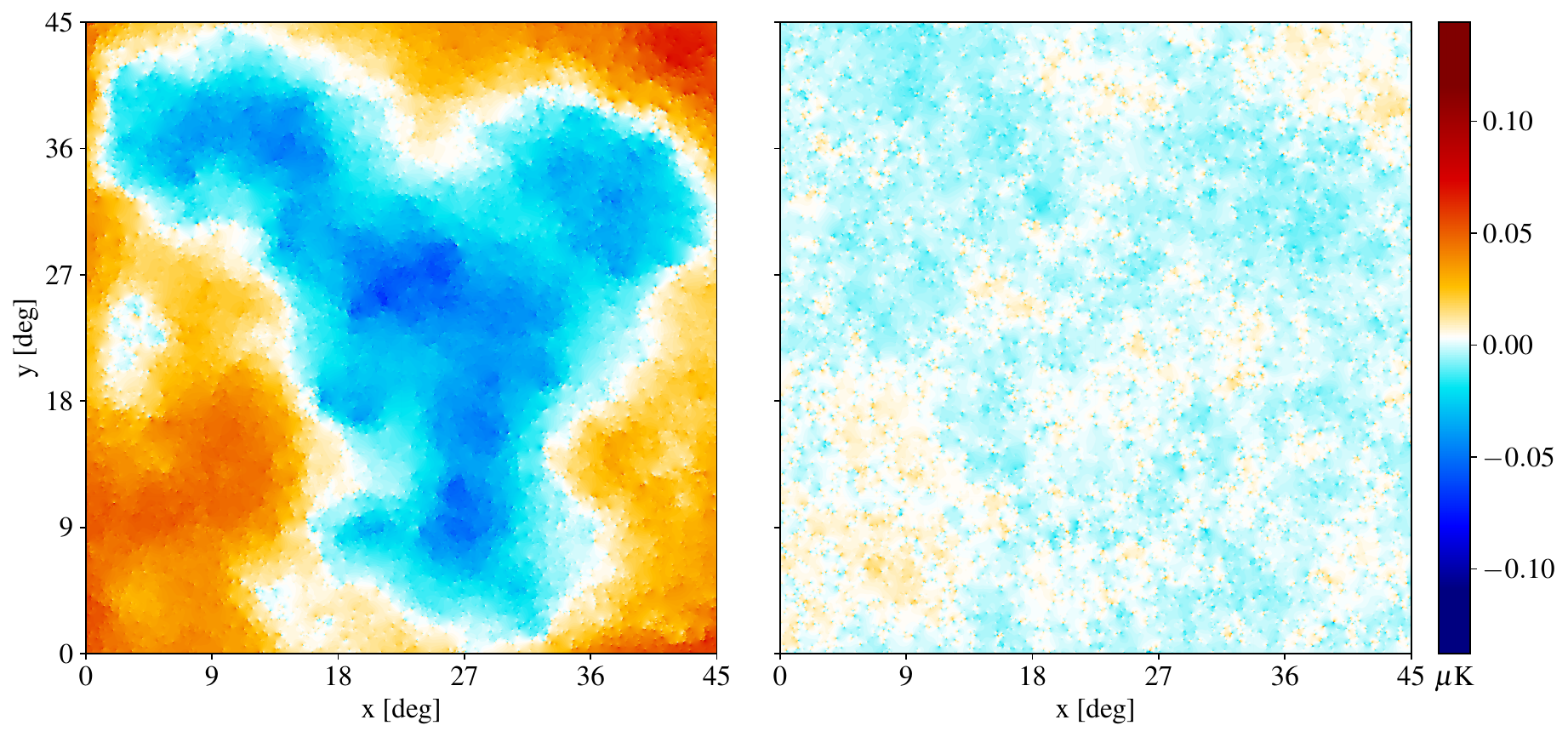}
    \caption{
    Simulated moving lens signal maps of the a catalog of galaxies with realistically correlated (left) and shuffled (right) velocities. 
    The shuffling of velocities removes the 2-halo term contribution to the average signal measured at each galaxy.
    }
    \label{fig:truVSsh}
\end{figure*}

Since the signal scales like $1/r$ from each halo center, the signal from nearby halos can add constructively or destructively depending on the relative direction of their velocities. 
For instance, if two nearby halos move in relatively the same transverse directions on the sky and they are located on the axis perpendicular to their transverse velocity direction separated by a few Mpc, their moving lens dipoles add constructively. But if the two haloes are both along an axis parallel to their transverse velocity, their moving lens dipoles will tend to cancel. 
In other words, the signal measured at the location of each halo can be written as the sum of two terms, one coming from its own moving lens signal (known as the 1-halo term) and one from the signals of other nearby halos around it (the 2-halo term). 
Importantly, the mean density of the Universe gives no moving lens effect, since a moving uniform distribution of mass does not lead to a time-varying gravitational potential. Therefore any visible imprint on the CMB from the moving lens effect has to come from excess mass density, and only the neighboring galaxies with correlated velocity contribute to the 2-halo term.
Since the velocity field of halos is coherent on large scales of around $100$~Mpc \cite{2023arXiv231212435R}), we expect to measure a significant 2-halo term in our stacking analysis, in particular at large aperture sizes. 
By shuffling the velocities of halos randomly across galaxies and simulating the shuffled moving lens map, we can null this 2-halo term.
Fig.~\ref{fig:truVSsh} shows two simulated maps of the a catalog of galaxies with true and shuffled velocities.
These two allow us to estimate the 1-halo and 2-halo terms separately.

\section{Model Observations}\label{pipeline}

\subsection{Aperture filter and stacked estimator}
\label{aper}

\begin{figure}[h]
    \centering
    \includegraphics[width=.27\textwidth]{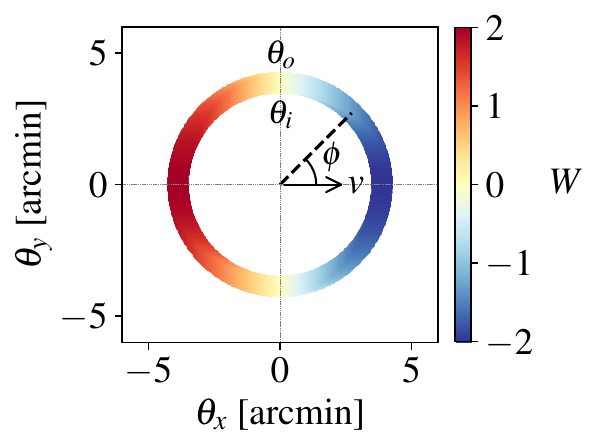}
    \caption{A cosine ring filter used to measure the moving lens signal. The vector $\bm{v}$ represents the velocity of galaxy, $\theta_i$ and $\theta_o$ are the inner and outer radius of the filter, and the angle $\phi$ is the angle between the velocity direction of the cluster and the radial line connecting the cluster to the points inside the ring aperture.
    }
    \label{fig:filter}
\end{figure}

\begin{figure}[h]
    \centering
    \includegraphics[width=.488\textwidth]{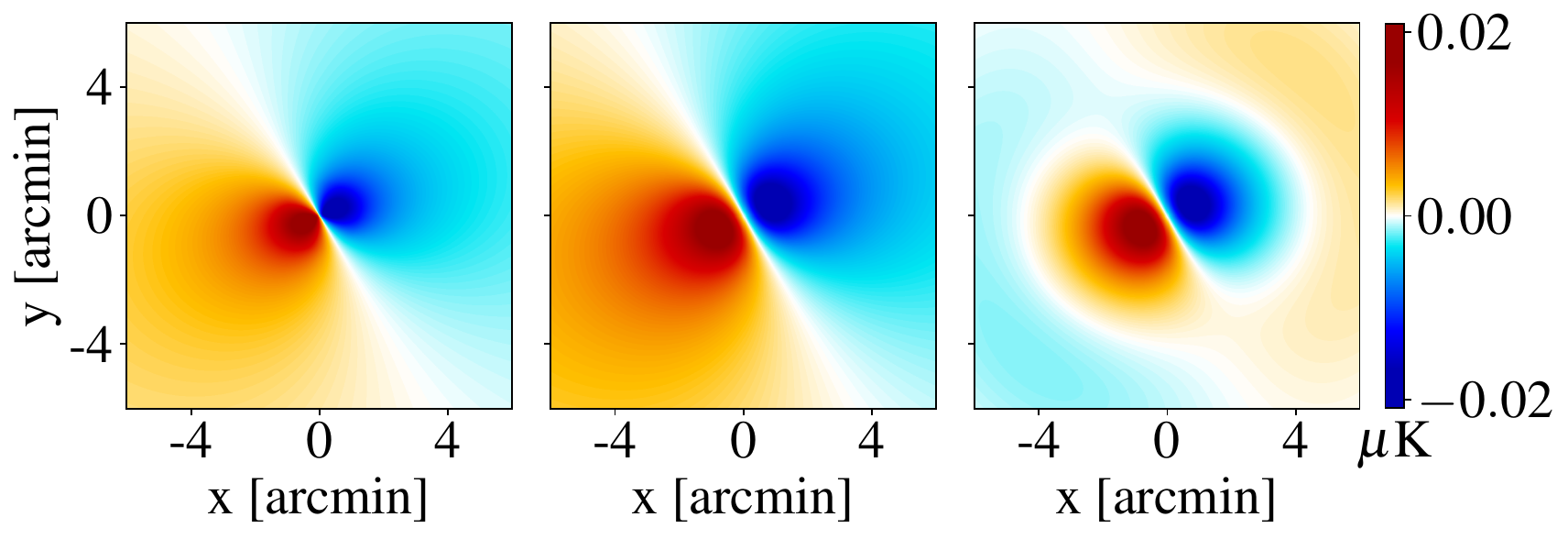}
    \caption{Dipolar pattern of anisotropy caused by a moving cluster (left), 1' beamed  (middle), 1' beamed and high-pass filtered with $\ell_\text{cut}=2000$ (right). This cluster  has a mass of $5 \times 10^{13}$ M$_\odot$, transverse velocity of $450$ km/s and redshift $z = 0.5$. 
    }
    \label{fig:dipole}
\end{figure}

In order to measure the moving lens signal of each cluster, we use concentric cosine ring aperture filters with varying inner radius $\theta_i$ and outer radius $\theta_o=\theta_i+0.625'$. 
We can write the output of the filter on the temperature map $\delta T$ as
\beq
\mathcal{T}(\theta_{i},\theta_{o})=\frac{1}{A}\int d\Omega \hspace{3pt} \delta T(\theta,\phi) \hspace{3pt} W_{\theta_{i},\theta_{o}}(\theta,\phi),
\label{eq:ap_filter}
\eeq
where $A$ is the angular area inside the ring and the filter $W_{\theta_{i},\theta_{o}}(\theta,\phi)$ is defined as
\beq
\begin{split}
    W_{\theta_{i},\theta_{o}}(\theta,\phi) =    
    \begin{cases}
         -2\cos(\phi), & \theta_{i}<\theta<\theta_{o} \\ 
         0, & \text{otherwise}
    \end{cases}
\end{split}
\label{eq:cosring_filter}
\eeq
such that the angle $\phi$ is the angle between the velocity direction of the cluster and the radial line connecting the cluster to the points inside the ring aperture, and the factor 2 is used to cancel out the $1/2$ that is introduced by the integration. Fig.~\ref{fig:filter} show an example of the cosine ring filter aligned with the velocity vector of a galaxy.
The ring filter is preferred over disk filters for our analysis because it reduces the correlation between apertures at different sky locations. 
Furthermore, we apply a high-pass filter to the maps to suppress large-scale noise correlations from the CMB primary anisotropies. 
Such a filter modifies the signal as it causes map-space ringing (which can be modeled), but it does not change appreciably the overall signal to noise ratio of the signal since most of it is obtained from the smaller apertures. 
Fig.~\ref{fig:dipole} shows the effect of beam convolution and a high pass filter with $\ell_\text{cut}=2000$ applied to the dipolar pattern of the signal in a cutout whose side size is equal to the diameter of the larger aperture filter that we use in this study. The high pass filter is a sigmoid (logistic) function in Fourier space of the form $1/(1+e^{-(\ell-\ell_\text{cut})/50})$.
We discuss the noise correlations and the effect of the high pass-filter further in section \ref{noise}.
For our stacking analysis, since the moving-lens signal is proportional to the transverse velocity of the clusters, we use a minimum-variance unbiased linear estimator that takes the velocity weighted mean of the temperature measured $\mathcal{T}_j(\theta_{i},\theta_{o})$ around each cluster $j$. 
This estimator can be written as:
\beq 
\label{vw}
\hat{T}_{\text{ML}}(\theta_{i},\theta_{o}) = 
\frac{1}{r_{v_\perp}}
\frac{v_{\perp,\text{rms}}^{\text{rec}}}{c} 
\frac{\sum_j^N \mathcal{T}_j(\theta_{i},\theta_{o})(v_{\perp,j}^{\textcolor{red}{\text{rec}}}/c)}
{\sum_j^N (v_{\perp,j}^{\textcolor{red}{\text{rec}}}/c)^2}
\eeq
where $v_{\perp,j}^{\textcolor{red}{\text{rec}}}$ is the transverse velocity of the galaxy cluster $j$,
and $r_{v_\perp} \equiv <v_{\perp}^{\text{rec}} v_{\perp}^{\text{true}}> / \sigma_{v^{\text{rec}}_{\perp}} \sigma_{v^{\text{rec}}_{\perp}}$ is the correlation coefficient between true and reconstructed velocity components across the line of sight, estimated using simulations.
In what follows, we set 
$r_{v_\perp} = 0.8$ \cite{2023arXiv231212435R,2023arXiv231212434H}, 
and discuss the impact of 
$r_{v_\perp} < 1$ on the signal-to-noise ratio in Sec.~\ref{systematics}.
The weightings and the prefactor $v_{\perp,\text{rms}}^{\text{rec}}/c  r_{v_\perp}$ are chosen such that this estimator provides an estimate for the typical size of the moving lens signal in units of $\mu$K (independent of $r_{v_\perp}$).
Indeed, the expectation value of our estimator (i.e., mean over realizations of the moving lens signal and map noise) is:
\beq
\langle \hat{T}_{\text{ML}}(\theta_{i},\theta_{o}) \rangle
=
\langle \beta \left( \theta_i, \theta_o \right) \rangle
\frac{v_{\perp,\text{ rms}}^\text{true}}{c},
\label{eq:estimator_mean}
\eeq
where
$\beta \left( \theta_i, \theta_o \right)$
is the modulus of the lensing deflection (Eq.~\eqref{eq:lensing_deflection}), integrated over the aperture Eq.~\eqref{eq:ap_filter}.
In App. \ref{app:vwuwCompar} we provide a derivation of this estimator \textcolor{violet}{and show that such a velocity weighted estimator will increase the signal to noise ratio of the moving lens effect by a factor of $2/\sqrt{\pi}\simeq 1.13$ (for perfect reconstructed velocities) compared to a uniform weighted estimator.}\\

Using the velocity-weighted estimator, we perform the stacking analysis on simulated maps of clusters with true and shuffled velocities based on \textcolor{red}{the constant mass (CMASS) catalog from BOSS} and obtain the 1-halo and 2-halo contributions to the moving lens signal profile as a function of aperture size $\theta_{o}$.
The result is presented in Fig.~\ref{fig:1-2-halo bhpf} below.
Interestingly, the 2-halo term is a negligible contribution to the signal, once the moving lens map is convolved with the beam and high-pass filtered.
In App. \ref{validation}, we discuss the more complex case where no beam or high-pass filter is applied, finding larger results for the 2-halo term.
\begin{figure}[h]
    \centering
    \includegraphics[width=.48\textwidth]{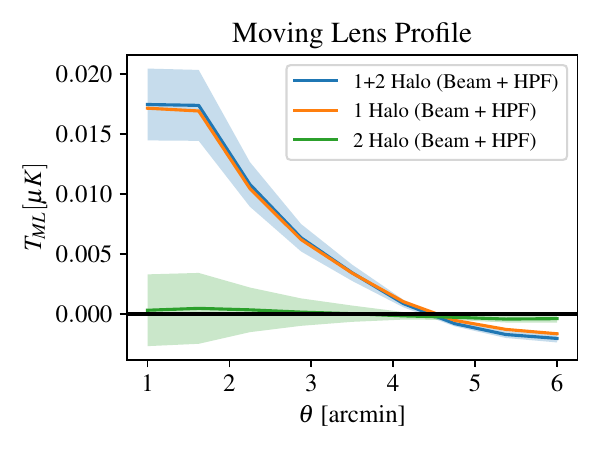}
    \caption{The stacked profiles of the 1-halo and 2-halo contributions to the moving lens signal for a simulated 45$\times$45 deg$^2$ patch of galaxy clusters based on BOSS catalog, after applying a high pass filter and 1 arcmin beam. 
    The curves include a correction at small apertures due to imperfect pixelation (see App.~\ref{validation}).
    }
    \label{fig:1-2-halo bhpf}
\end{figure}

\subsection{CMB experiments and LSS surveys}\label{assump}

We utilize the CMASS catalog from BOSS DR12, which incorporates reconstructed velocities, as our primary galaxy catalog for simulating the moving lens signals. 
Specifically, we use the reconstructed velocities from \cite{2017MNRAS.467.2331V,2016PhRvD..93h2002S} to simulate the true velocity correlation across galaxies.
Our analysis extends beyond CMASS to encompass additional galaxy surveys, including DESI, unWISE, and LSST. 
The respective galaxy counts for these surveys are detailed in Table \ref{snr_table} in Section \ref{noise}. For the DESI survey, we adopt surface density values of 602 and 1677 galaxies deg$^{-2}$ for its luminous red galaxies (LRGs) and extended LRGs, respectively~\cite{2023JCAP...11..097Z}. 
Similarly, we assume surface densities of 5400 galaxies deg$^{-2}$ for unWISE~\cite{2022PhRvD.106l3517K} and 48 galaxies arcmin$^{-2}$ from LSST's gold sample~\cite{2018arXiv180901669T}, contributing to the comprehensiveness of our estimates.
Additionally, we use the Data Challenge 2 (DC2) extragalactic sample \cite{dc2} to account for LSST's broad mass distribution of halos as well as finding its mean halo mass ($6.4\times 10^{12} M_\odot$).

\begin{figure}[h]
    \centering
    \includegraphics[width=.48\textwidth]{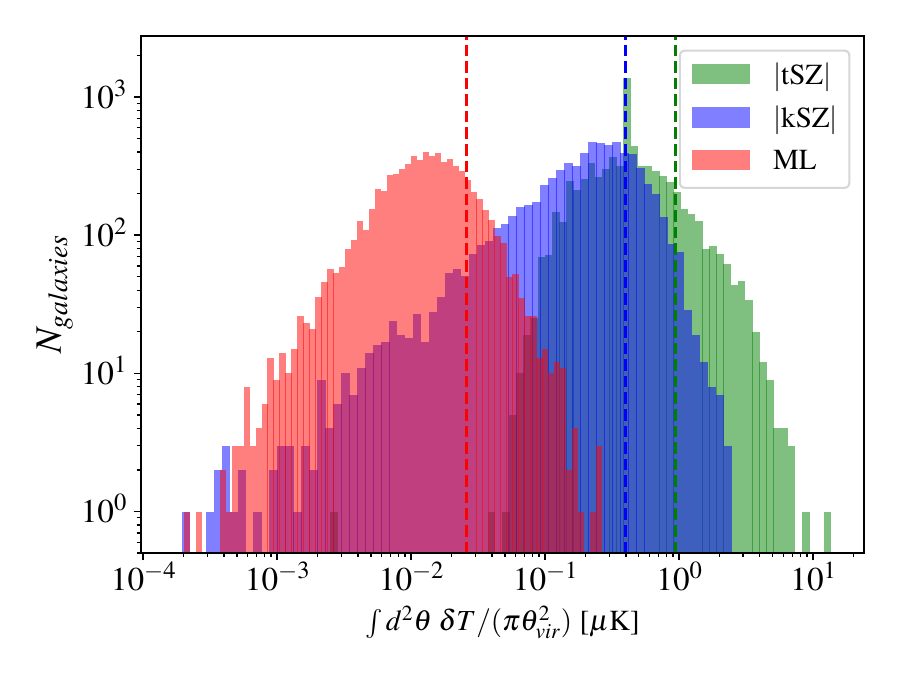}\\
    \caption{Histograms of estimated signals for the clusters in a 10$\times$10 deg$^2$ patch of the BOSS catalog. The tSZ effect is calculated at 150 GHz frequency. The dashed lines represent the RMS values of the tSZ, kSZ and ML signals (0.938, 0.397 and 0.026 $\mu$K respectively). The spike in tSZ corresponds to the mean mass of the sample and represents galaxies for which we did not have a mass estimate.
    For the moving lens whose effect is a dipole, the aperture integral also includes a dipole weight ($-2\ \text{cos}\ \phi$) to avoid cancellation (Eq.~\eqref{eq:cosring_filter}).
    }
    \label{fig:histograms}
\end{figure}

To gain insight into the relative magnitudes of the moving lens signal compared to the thermal and kinetic Sunyaev-Zeldovich effects, we present a histogram in Fig.~\ref{fig:histograms}. 
This histogram illustrates the integrated signals obtained from a 10$\times$10 deg$^2$ region of the CMASS catalog of BOSS. In these calculations, we assume all baryons within the halo are in the form of homogeneous fully ionized gas, and the total baryon mass is derived from the \textcolor{red}{virial} mass of the halo using the cosmological baryon fraction $\Omega_b / \Omega_m$.
In our analysis of CMB experiments, we consider the Atacama Cosmology Telescope, Simons Observatory and CMB Stage-IV, characterized by beam sizes of 1.4, 1.4, and 1 arcmin, and sensitivity levels of 10, 5, and 1 $\mu$K-arcmin, respectively. 
Our forecasts involve a sky overlap between the DESI survey and the CMB experiments encompassing 1500 deg$^2$ in the first year (Y1) and expanding to 5000 deg$^2$ in the fifth year (Y5). 
Additionally, for unWISE and LSST, we assume complete coverage with the ACT experiment, totaling approximately 9400 deg$^2$.

\section{Simulated stacking detection}
\label{results}

\subsection{Noise simulations and forecasts}
\label{noise}

\begin{figure}[h]
    \centering
    \includegraphics[width=.49\textwidth]{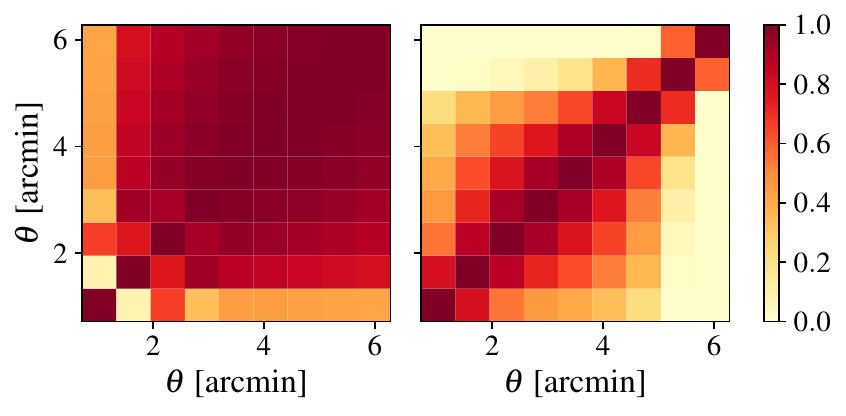}
    \caption{Correlation matrices of the CMB-S4 experiment's noise (lensed CMB, detector and atmospheric noise, extragalactic foregrounds at 150~GHz) between the different cosine ring apertures in the absence of high pass filtering (left) and after applying a high-pass filter with $\ell_\text{cut}=2000$ (right).}
    \label{fig:corMat cosring no hpf}
\end{figure}

In the forecasts below, we perform bootstrap resampling of the galaxy catalog and repeat our stacking measurement to estimate the covariance matrix of the measured profile.
Bootstrap resampling assumes that the various ring filters around each galaxy \textcolor{red}{are} independent.
This assumption breaks down for the densest galaxy surveys considered here (unWISE and LSST), as shown in \cite{2024arXiv240113040S}.
\textcolor{red}{In these cases, a more accurate covariance could be obtained by resampling in sufficiently large spatial blocks or repeating the analysis on many simulations which would be computationally expensive; we leave this to future work.}
In practice, this is at most a factor of $\simeq$2.5 overestimate of the SNR for LSST.
For unWISE, this is a 20\% change only, and we expect a much smaller effect for DESI and the lower density samples (see Table~III
 in~\cite{2024arXiv240113040S}).
Fig.~\ref{fig:corMat cosring no hpf} shows
how the high-pass filtering greatly reduces the correlation coefficient across apertures (right) compared to a raw temperature map (left).
This makes interpreting the observed profiles easier, by decorrelating the various aperture radii.
We considered several high-pass filters with different $\ell_\text{cut}$ for this purpose, and decided to use the $\ell_\text{cut}=2000$ as it could provide a significant decorrelation between apertures while almost preserving the profile of the signal despite the ringings. 
In App. \ref{validation}, we show the impact of different high pass filters with varying $\ell_\mathrm{\textcolor{red}{cut}}$ on the profile of the moving lens signal.  

For the ACT experiment we have modeled the noise based on the noise power spectrum of ACT DR6 \textcolor{red}{Internal Linear Combination} (ILC) map, while for SO we used the forecasted ``goal'' component separated noise power spectrum \cite{SOScience}. 
For simulating the noise of S4 experiment, we simulated a single frequency noise map at 150 GHz containing lensed CMB, detector and atmospheric noise,  as well as extra-galactic foregrounds.

\begin{figure}[h]
    \centering
    \includegraphics[width=.45\textwidth]{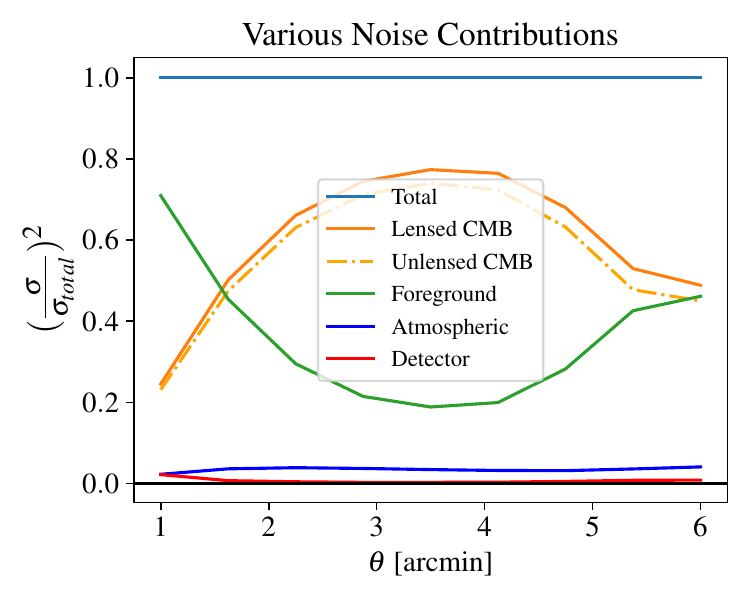}\\
    \caption{Standard deviations of different noise sources for CMB-S4 experiment obtained from \textcolor{red}{high-pass} filtered simulated \textcolor{red}{sky maps without the moving lens effect}.
    The various contributions add in quadrature. The dashed line is a hypothetical case that shows how the CMB noise contribution will change if we use the unlensed power spectrum to generate that component (see Sec.~\ref{lensing}).
    }
    \label{fig:S4_noises}
\end{figure}

To have a better understanding of the contribution of different sources of noise to our measurements we show their individual estimated standard deviations as a function of cosine ring apertures in Fig.~\ref{fig:S4_noises}. 
This plot shows that in the smallest apertures the \textcolor{red}{foreground} noise is the dominant component while as we go to larger apertures the lensed CMB contributes the most to the total noise. 
\textcolor{red}{As mentioned earlier, our noise simulations include detector and atmospheric noise, lensed CMB and foregrounds, but not the moving lens signal itself. There are two reasons for this. First, since the moving lens effect has not yet been detected, we are focusing on forecasting the detection significance, rather than the precision on the amplitude of the effect. The difference is that when assessing the detection significance, the question we ask is how well we can rule out the null hypothesis, i.e. the hypothesis of no moving lens signal. In this null hypothesis, one therefore should not include the signal to be detected, i.e. one should not include the cosmic variance of the moving lens signal. This point of principle is true in general: when ruling out the hypothesis of no signal, one performs an analysis without the signal of interest and shows that the resulting model is a poor fit to the data. Second, once the moving lens signal is detected, and we move from ruling out the “no moving lens” hypothesis to measuring the amplitude of this signal, the situation changes: one should include the cosmic variance of the signal. In other words, one should include the moving lens signal in the “noise” simulations. However, this cosmic variance would be negligible anyway. Indeed, in a stacked analysis, the cosmic variance of the signal matters once the signal is comparable in size with the other sources of noise on an individual object basis. That is, the cosmic variance starts to matter once the SNR of the moving lens effect on an individual object is of order unity. For all the upcoming experiments considered in our forecast, we are very far from detecting the moving lens effect per object, such that this cosmic variance is negligible.}

Finally, we present the signal to noise ratio forecast for various combinations of the CMB experiments and galaxy surveys in Table~\ref{snr_table}. 
These SNR values are 
calculated using the vector $\text{S}$ containing the amplitude of the stacked moving lens signal at different apertures and the noise covariance matrix $\text{C}_N$ between apertures obtained from the simulated \textcolor{red}{sky maps without the moving lens effect} via bootstrap, as follows:
\beq
\text{SNR}^2 = \text{S}^T \cdot \text{C}^{-1}_N \cdot \text{S}
\eeq
This number is indeed the SNR on an amplitude factor $a$ one would fit for, if the template for the spatial dependence of the moving lens signal is assumed to be known.

\textcolor{red}{We consider various variations of the ILC method, comparing the standard ILC (the multifrequency linear combination which maximizes the overall SNR), and ILCs which null tSZ or Cosmic Infrared Background (CIB). 
In this table, the values in parentheses correspond to further deprojecting tSZ, CIB or both in the internal linear combination of recovering CMB maps. 
Table~\ref{snr_table} shows that the tSZ or CIB deprojected ILCs cause such a large SNR penalty that a detection of the moving lens effect would be impossible with these methods. We thus focus on the standard ILC in the rest of the paper. We assess the impact of the residual tSZ and CIB signals and their potential biases in the moving lens measurements in Sec.~\ref{biases}, where we present and implement a mitigation strategy on single frequency maps, selecting frequencies which maximize the impact of tSZ and CIB, in order to provide an upper limit to the size of the foreground biases and show that \textcolor{violet}{for CMASS like samples} our mitigation strategy is sufficient in this worst case scenario, giving confidence that it should apply in the case of the standard ILC, where the foreground amplitude is less than in these single frequency maps. \textcolor{violet}{We will also mention in Sec.~\ref{biases} that for samples with higher mass and redshift, the foreground biases are more of a concern and further mitigation strategies will be required.}}
\begin{table*}[htbp]
\resizebox{\textwidth}{!}{
\setlength{\tabcolsep}{12pt} 
\begin{tabular}{ c c c c c }
\hline \\
\textbf{SNR}             & \textbf{\#Galaxies overlapping}    & \textbf{ACT}             & \textbf{SO}                & \textbf{S4}    \\ \\ \hline
\textbf{BOSS}                   & 350k       &  1.2  (0.2)    &  1.4 (0.7, 1.4, 0.3)        & 1.8   \\ \hline
\textbf{DESI LRGs Y1}           & 900k       &  1.8  (0.4)    &  2.3 (1.2, 2.2, 0.5)    & 2.8   \\ 
\textbf{DESI LRGs Y5}           & 3M         &  3.4  (0.7)    &  4.3 (2.2, 4, 0.9)    & 5.1   \\ 
\textbf{DESI LRGs + Ext. LRGs Y1} & 3.4M       &  3.7  (0.8)    &  4.6 (2.3, 3.4, 1)    & 5.4   \\ 
\textbf{DESI LRGs + Ext. LRGs Y5} & 11.4M      &  6.4  (1.4)    &  8.3 (4.2, 7.8, 1.8)    & 9.9  \\ \hline
\textbf{unWISE}                & 50.8M      &  11.9  (2.6)   &  14.8 (7.6, 14.0, 3.1)  & 17.8 \\ \hline
\textbf{LSST} equal weight (pessimistic)                  & 1.12B      &  10.2 (2.2) & 12.7 (6.5, 11.9, 2.7) & 15.1 \\ 
\textbf{LSST} equal weight; $10^{12} $M$_\odot <$M                  & 330M      &  18.0 (4.0) & 22.5 (11.4, 21.1, 4.7) & 26.8 \\ 
\textbf{LSST} equal weight, $10^{14} $M$_\odot <$M                  & 8.4M      &  51.6 (11.4) & 64.6 (32.8, 60.4, 13.4) & 76.7 \\ 
\textbf{LSST} mass weight (optimistic)                   & 1.12B      &  68.1  (15.1) &  85.3 (43.2, 79.8, 17.6) & 101.3 \\ \hline
\end{tabular}
}
\caption{
Moving lens detection signal-to-noise forecasts for different galaxy surveys and CMB experiments. The SNR values in parentheses for ACT correspond to deprojecting tSZ in the ILC, and for SO to deprojecting \textcolor{red}{tSZ}, CIB, or both in the ILC. 
Realistic velocity reconstruction results in a factor 0.8 degradation in SNR~\cite{2023arXiv231212435R} which is included in this table. We discuss the impact of velocity reconstruction and the effect of miscentering between galaxy positions and center of host halo on the signal to noise ratios in  Sec.~\ref{systematics}.
The LSST forecast is corrected by a factor 2 \cite{2024arXiv240113040S} since the bootstrap method of noise estimation is optimistic as it neglects the correlations of nearby galaxies. For unWISE, this is roughly a 15$\%$ change in SNR which is accounted for in the table, and for lower density galaxy samples the difference is negligible.
For LSST, the broad range of halo masses makes the equal weighting highly suboptimal, compared to mass weighting (which would optimistically require individual mass estimates).
In practice, a low mass cut can likely be implemented based on galaxy brightness. \textcolor{violet}{Imperfect foreground cleaning at higher redshifts may incur a SNR penalty; see the discussion in Sec.~\ref{noise} and Sec.~\ref{biases} on foregrounds.}
}
\label{snr_table}
\end{table*}
In practice, the imperfect reconstructed velocities can lead to a degradation in SNR by a factor equal to $r_{v_\perp}\simeq 0.8$.
This is discussed in more detail in Sec.~\ref{systematics} and Ref.~\cite{2023arXiv231212435R} and accounted for in Table~\ref{snr_table}. 

For the LSST survey, which encompasses a broad range of galaxy masses, we provide four forecasts for each CMB experiment. 
The first forecast assumes no \textcolor{red}{individual} mass information, assigning equal weight to all galaxies, resulting in the lowest signal-to-noise ratio. The second and third forecasts also employ equal weighting but apply mass cuts of $10^{12} \, M_\odot < M$ and $10^{14} \, M_\odot < M$, respectively, based on the mass distribution from the DC2 extra-galactic sample \cite{dc2}. The final forecast incorporates knowledge of individual galaxy masses, weighting the signal contribution according to these masses, and achieves the highest SNR. Due to the relatively low mean mass of the overall sample (approximately $6.4 \times 10^{12} \, \text{M}_\odot$), there is a notable improvement in SNR when mass information is utilized and galaxies are weighted accordingly in the estimator. Importantly, the SNR is predominantly influenced by the more massive galaxy clusters, around $10^{14} \, \text{M}_\odot$. However, at these high masses, the foreground emission and halo lensing associated with these objects may be substantially greater compared to random sky regions, potentially leading to an overestimation of the SNR by approximately tens of percent. We leave further investigation of this to future work.

In conclusion, the results of Table~\ref{snr_table} suggest that a high significance detection of the moving lens effect should be achievable in the near future with the combination of data from SO/S4 \textcolor{red}{experiments} and LSST survey. Additionaly, the forecast for the latest ACT$\times$DESI data shows a promising prospect for finding evidence of this effect. 
We also note that improvements in CMB experiment sensitivity and resolution do not significantly enhance the moving lens detection SNR, whereas increasing the number of galaxies in the LSS survey can substantially improve the SNR. This is consistent with the fact that the noise in this measurement is primarily dominated by other sky signals 
rather than detector noise, as shown in Fig.~\ref{fig:S4_noises}.

\subsection{Consistency with existing forecasts}
\label{sec:comparison_hotinli}

We compare our forecasted SNR to the results of Ref.~\cite{HotPaol}.
That study reports a signal-to-noise ratio of $\simeq$21 for the LSST$\times$S4 combination, assuming that the signal around each galaxy is equally weighted, with no mass cuts on the sample beyond the LSST selection.
For such a scenario, we would find SNR of $\simeq$28, slightly higher, perhaps due to our neglect of the noise from halo lensing which may contribute a non-negligible noise for halos of masses $\gtrsim 10^{14} M_\odot$.

However, for a galaxy catalog with such a high number density, the \textcolor{red}{jacknife} (\cite{HotPaol}) or \textcolor{red}{bootstrap} (this work) estimates of the noise are underestimated.
Indeed, both pretend that the temperatures measured at the positions of any two objects are independent.
This assumption, valid when the galaxy catalog is shot noise dominated, breaks down at high galaxy number density.
To account for this effect, we use the SNR reduction factor found in \cite{2024arXiv240113040S} and divide our SNR by it, reducing it 15.
This same SNR reduction factor is included in all the LSST forecasts in this paper.

Furthermore, the LSST sample considered contains a wide breadth of halo masses.
The galaxy count is dominated by the abundant low-mass halos, which contribute almost no signal and dilute the stack. 
In this context, the equal-weighting of object thus becomes highly suboptimal.
If individual host halo masses are known, mass weighting enhances the SNR by a large factor of $\simeq$6. 
Such an ideal mass weighting may not be possible in practice; we can however approach it with a simple low mass cut, perhaps based on galaxy brightnesses.

The other differences in assumptions between this work and Ref.~\cite{HotPaol} should have a small impact on the forecast.
Their study assumes lower resolutions and sensitivities for S4, though this is not a significant factor, as shown in Fig.~\ref{fig:S4_noises}, where we demonstrate that the analysis is not dominated by detector noise. The total number of galaxies assumed in both studies is nearly identical, with our work assuming a higher galaxy surface density but a lower sky fraction overlap.
Finally, in our simulations we have considered a truncation of the NFW density profile at $R_{\text{vir}}$ radius whereas they applied an exponential truncation of the moving lens profile at $8R_{200c}$ using \texttt{AstroPaint}. While this can affect the 2-halo term contribution to the signal we showed in Fig.~\ref{fig:1-2-halo bhpf} that such contribution is negligible after the high-pass filtering of the map.

\subsection{Foreground biases and mitigation strategies}\label{biases}

The cosine-ring aperture we apply around each galaxy should null the dust and tSZ signals from that particular galaxy.
Indeed, this dipole filter integrates to zero, and the dust and tSZ signals from the galaxy considered have no angular dependence on average.
However, the dust and tSZ signals from other objects may bias our moving lens estimate \cite{2023PhRvD.108h3508H}.
Indeed, the cosine-ring filter is oriented in the direction of the peculiar velocity of the galaxy considered.
The large-scale linear peculiar velocity field preferentially points towards overdense regions, which cause the peculiar motion in the first place, and whose dust and tSZ signals are higher.
Thus we expect the dust and tSZ maps around each galaxy to have a dipole structure aligned with the peculiar velocity, leading to a bias in the cosine-ring filter.

\begin{figure*}[htp!]
    \centering
    \includegraphics[width=.85\textwidth]{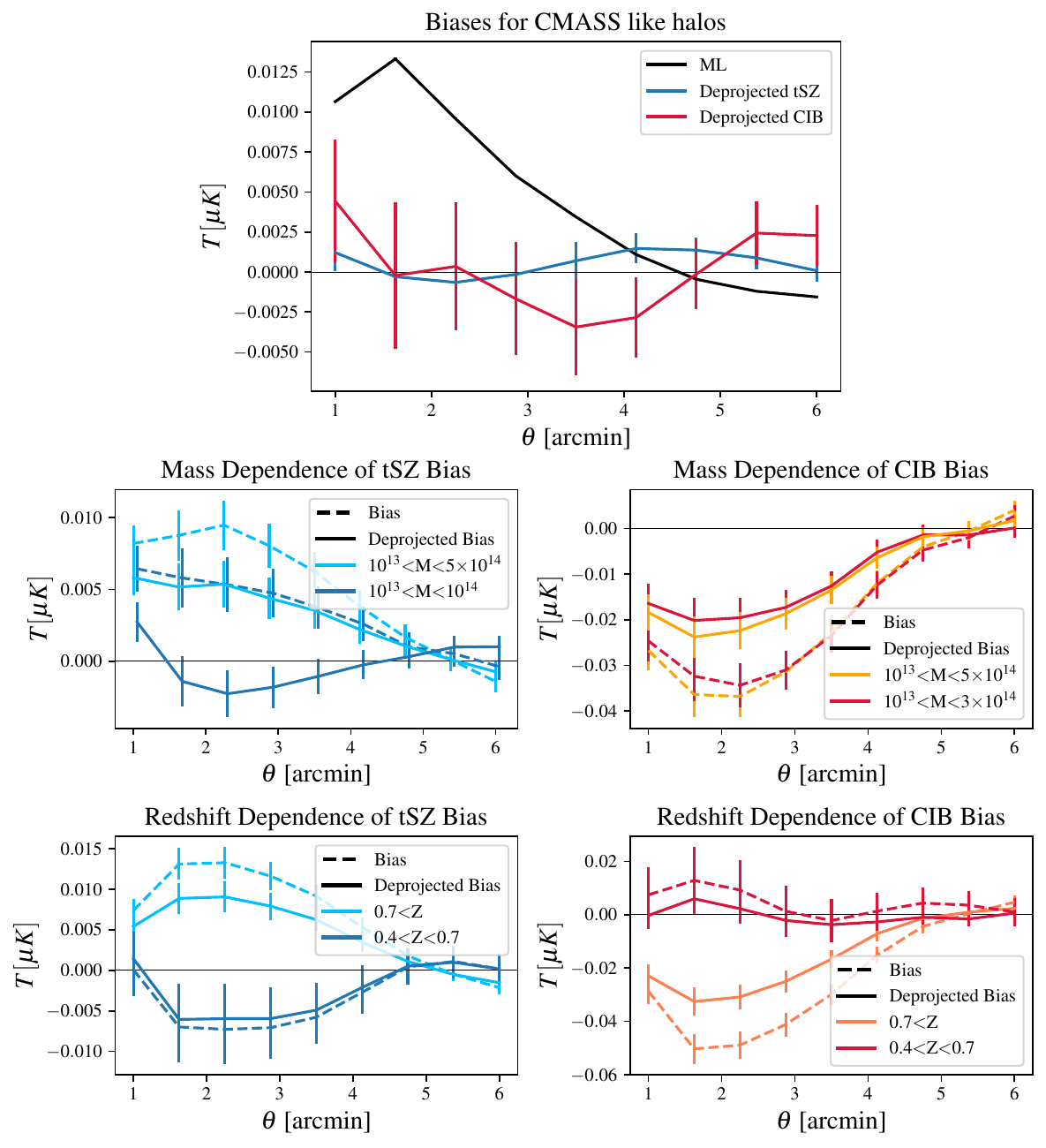}\\
    \caption{\textbf{Top.} The stacked profiles of the moving lens signal for a sample of galaxies selected from BOSS catalog in comparison to the stacked profiles of the deprojected tSZ and CIB biases obtained from a sample of halos in \texttt{WebSky}. We can see that the template deprojection method along with masking of the most massive objects can reduce the sizes of these biases to few factors smaller than the moving lens signal. \textbf{Bottom.} The 4 panels below are showing the mass and redshift dependence of tSZ and CIB biases. The dashes lines represent the profiles of the biases prior to the cleaning method and the solid lines show the results after performing the template deprojection. As described in the text, the samples that include higher redshift or higher mass clusters introduce larger biases, however for lower masses and redshifts the amplitude of the biases \textcolor{red}{after the template deprojection can be reduced down to 2-5 times smaller than the moving lens signal, especially at low redshifts.}
    }
    \label{fig:biases_all}
\end{figure*}

In this subsection, we confirm the existence of these spurious signals in our moving lens measurements by running our pipeline on 20$\times$20 deg$^2$ patches extracted from CIB and tSZ \texttt{WebSky} maps \cite{2019MNRAS.483.2236S,2020JCAP...10..012S}. 
We select stacking samples with different redshift and mass ranges from the \texttt{WebSky} halo catalog to investigate the redshift and mass dependence of the biases. 
In agreement with Ref.~\cite{2023PhRvD.108h3508H}, we find in Fig.~\ref{fig:biases_all} that for high mass or high redshift samples these biases are indeed significant, so a mitigation strategy will be needed. 
However, if we only consider relatively low redshift and masses such as the galaxies in the CMASS catalog of BOSS for which we investigated the moving lens signal in previous sections, the tSZ and CIB biases are not as troublesome (see Fig.~\ref{fig:biases_all}).

Furthermore, several foreground mitigation approaches exist to reduce this foreground bias to an acceptable level.
Dust emission and tSZ decrement partially cancel in single frequency maps at frequencies below 220~GHz, since they have opposite signs.
Multifrequency cleaning reduces the amplitude of \textcolor{red}{dust or tSZ} in our maps.
However, as we can see in Table~\ref{snr_table}, such deprojections boost the overall CMB noise to an extent where the SNR gets degraded significantly, so an ILC with tSZ/dust deprojection is not preferred over the standard ILC for the purpose of the moving lens detection.
Masking the individually-detected point sources and clusters in the map is also an effective way to remove the high-mass objects, whose foreground bias is largest.

Finally, we propose and demonstrate an additional foreground mitigation technique, based on template deprojection.
The dust and tSZ signals from the galaxies in our sample are much larger than their moving lens effect.
They can thus be measured at much higher SNR (more than 10 times higher) with a standard isotropic stack (e.g., ring filters without the cosine angular dependence) or by cross-correlating the temperature map with the galaxy catalog.
One can then generate a foreground template map where each galaxy is given the dust + tSZ signal thus measured.
Running the cosine ring filter on this template map should then capture part of the foreground bias to the moving lens estimator.
Indeed, this template captures the effect that peculiar velocities point towards overdense regions, which have more galaxies, and thus more foreground emission.
However, the foreground template paints the same signal around each galaxy in the catalog.
Thus, it does not capture the effect that overdense regions may host more massive halos, whose individual foreground emissions would also be higher.
For the same reason, the template deprojection will also be ineffective when the foreground bias is dominated by a few rare bright objects.
This can be avoided by masking the brightest sources in the map before deprojection.

In practice, the simplest way to implement this mitigation method is via template deprojection (see \cite{2023PhRvD.108l3501K} for a similar idea).
We can think of the temperature map $T_\vl$ in Fourier space as the sum of a contribution correlated with 2D galaxy number overdensity field $g_\vl$ and a contribution $T^\text{clean}_\vl$ that is uncorrelated with $g_\vl$ such that:
\beq
T_\vl 
= 
\alpha_\ell g_\vl
+
T^\text{clean}_\vl
\quad\text{with}\quad
C^{g T^\text{clean}}_\ell = 0.
\eeq
Now if we choose $\alpha_\ell = C_\ell^{gT} / C_\ell^{gg}$ we will have the cleaned temperature map in Fourier space as
\beq
T^\text{clean}_\vl
=
T_\vl
-
\frac{C_\ell^{gT}}{C_\ell^{gg}}
g_\vl.
\eeq
In this cleaned map, the quantity 
$(C_\ell^{gT} / C_\ell^{gg})
g_\vl$
can be thought of as a foreground template in real space, where the same foreground profile
$(C_\ell^{gT} / C_\ell^{gg})$
is painted around each object.
The cleaning does not remove any of the moving lens effect, since this signal already satisfies 
$C^{ g \delta T_\text{ML}}_\ell = 0$.

Fig.~\ref{fig:biases_all} shows the effectiveness of masking and the template deprojection on the realistic \texttt{WebSky} tSZ and CIB patches.
The top panel shows the profiles of the tSZ and CIB biases after performing the mitigation for a sample of CMASS like halos in the \texttt{WebSky} catalog, in comparison to the stacked moving lens signal profile.
The mitigated foreground bias is consistent with zero, and smaller than the moving lens signal by a factor of several.
The bottom panels show the mass and redshift dependence of the biases, as well as the effect of the mitigation strategy. 
In each panel, the dashed lines represent the bias profiles prior to the cleaning process and the solid lines show the profiles after performing the deprojection strategy. 
\textcolor{violet}{As expected, the samples that include higher masses or redshifts introduce a larger bias. In such cases, this cleaning method may prove insufficient, and additional mitigation strategies may be required, which we leave for future work.}
The template deprojection cleaning process performs better whenever the highest mass halos are not included in the map (e.g., after masking).
Indeed, since the template deprojection removes the same average dust and tSZ profile around each galaxy, it is not effective when a handful of rare bright objects dominate the bias. 

Finally, several null tests from the data are possible. 
For instance using sine filters rather than cosine nulls the moving lens and foreground biases, or doing the stacks on ILC maps where spectral energy distributions of tSZ/CIB/both are deprojected reduces or nulls one or more of the foreground biases. 
Performing the stack on a Planck map at 545 GHz could also allow us understand the dust component better and verify its effective mitigation.

\subsection{Impact of CMB lensing noise}\label{lensing}

CMB lensing due to the host halo of the galaxies in our catalogs produces local dipoles in the temperature map, similar to the moving lens effect.
In fact, the CMB lensing dipoles are much larger in amplitude than the moving lens effect, and have already been detected at high significance (see e.g., \cite{2015ApJ...806..247B,PhysRevLett.123.181301})
However, these dipoles are oriented along the unlensed CMB gradient, rather than the peculiar velocities.
Since these two vector fields are independent, CMB lensing does not bias our moving lens effect. 
One may worry that it constitutes a dominant source of noise, since it is much larger in amplitude.
However, we find this not to be the case. 

Indeed, in the absence of foregrounds or detector noise,
the CMB power spectrum is dominated by lensing for multipoles $\ell \gtrsim 5000$.
One may thus worry that the aggregate lensing signal from all massive objects in the Universe could be the dominant source of noise in a futuristic experiment.
We account for this effect by including the lensed CMB power spectrum in our simulated temperature maps.
In practice, due to the presence of extragalactic foregrounds, replacing the lensed CMB power spectrum by the unlensed CMB power spectrum makes little difference to the noise, as shown in Fig.~\ref{fig:S4_noises} by solid and dashed orange curves.

On the other hand, since the galaxies we consider are large non-Gaussian fluctuations in the matter density field, one may worry that their lensing may be a larger source of noise than included in the lensed CMB power spectrum.
However, if the lensing dipole from a typical galaxy in our sample was the dominant noise source, larger than \textcolor{red}{the} primary CMB, detector noise and residual foregrounds around that galaxy,
it would imply that the lensing signal from that galaxy can be detected individually.
In other words, CMB cluster lensing would be detectable around each galaxy individually.
This is not the case, as large galaxy catalogs are required to measure CMB lensing in cross-correlation.
\textcolor{red}{Indeed, Fig. 53 in the CMB-S4 Science Book \cite{2016arXiv161002743A} (similarly Fig. 5 in \cite{HuDeDeo}) shows that an arcminute resolution experiment with a sensitivity of around 1$\mu$K-arcmin can determine the mass of 1000 stacked clusters
(of mass $5\times 10^{14} M_\odot$) to approximately $2\%$ precision, combining temperature and polarization maps.
This means a halo lensing SNR of approximately $1/0.02 = 50$ is expected for stacking on 1000 objects of mass $5\times10^{14}M_\odot$. However, for the same number of objects but with a typical mass of $2\times10^{13}M_\odot$ in our sample, the expected SNR can be rescaled to $50\times(2\times10^{13}M_\odot/5\times10^{14}M_\odot)\simeq2$. Now, if we consider the SNR on a per object basis it will become $2\times(1/\sqrt{1000})\simeq0.06$,
which implies that halo lensing will be much smaller than other sources of noise, even for the optimal halo lensing estimator, and \textit{a fortiori} for the moving lens estimator.
Interestingly, this estimate suggests that halo lensing will become a non-negligible source of noise for more massive halos, $\sim 10^{14} M_\odot$, since the halo lensing signal scales roughly proportionally to halo mass.}

\subsection{Other systematics and potential biases}
\label{systematics}

\begin{figure}[h]
    \centering
    \includegraphics[width=.45\textwidth]{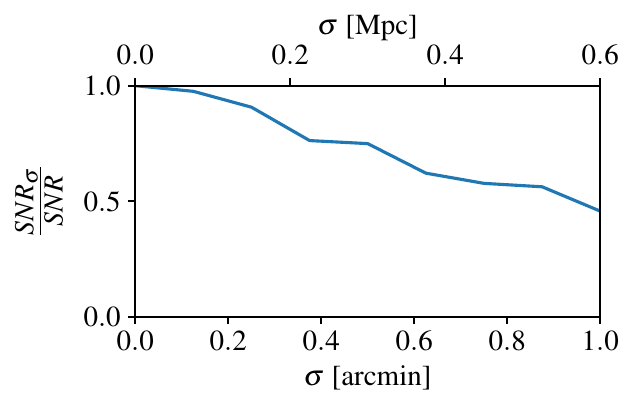}
    \caption{The impact of miscentering error on SNR forecasts assuming CMB-S4 experiment specifications.
    The miscentering between position of galaxies and halo centers galaxy is estimated to be $\simeq$ 0.2$'$ \cite{2012ApJ...757....2G}, hence we do not expect a large loss of SNR ($\simeq 5\%$) due to it.
    }
    \label{fig:miscentering}
\end{figure}

Several systematic effects may affect our measurement of the moving lens effect.
This can bias our interpretation of the signal, and may also lower the detection SNR, compared to our estimates above.
In this subsection, we discuss miscentering between the galaxy positions and the center of mass of the host halos\textcolor{red}{,} total matter profiles, satellite fraction in the galaxy sample, and systematic effects related to imperfect velocity reconstruction from the galaxy number density field.
To find out the impact of miscentering error of our filters on the signal to noise ratio, we add Gaussian noise of standard deviation $\sigma$ to the positions of halos on the sky and run our pipeline for each size of the miscentering noise. 
Fig.~\ref{fig:miscentering} shows the ratio of SNR with miscentered filters over SNR of perfectly centered filters as a function of miscentering noise size. 
The miscentering between the galaxy positions and halo centers is estimated to be $\simeq$ 0.2$'$ \cite{2012ApJ...757....2G}, hence Fig.~\ref{fig:miscentering} suggests that only a $\simeq$ 5$\%$ degradation in SNR can be expected due to the miscentering effect.
By broadening the observed matter density profile with respect to the true one, miscentering can bias our profile inference.
Unlike tSZ or kSZ, where the gas profile may be the quantity of interest, here the matter density profile is not.
However, it multiplies the amplitude of the transverse velocity in the moving lens effect, and may thus bias our inference of the transverse velocity amplitudes.

The satellite fraction of galaxies in the sample has two effects.
First, it introduces miscentering, discussed above.
Moreover, it may also affect the performance of the velocity reconstruction from the galaxy number density field.
Fortunately, Ref.~\cite{2023arXiv231212435R} shows that the satellite fraction has little effect on the transverse velocity reconstruction.
This can be understood for two reasons.
First, while satellite galaxies typically have much larger peculiar velocities due to virial motion inside their host halos, this large virial velocity is not that of the host halo, which is relevant for the moving lens effect.
Furthermore, the velocity reconstruction process is actually shown in \cite{2023arXiv231212435R} to reconstruct the halo velocities, rather than the galaxy velocities.
Indeed, the velocity reconstruction only considers large-scale linear motions, and does not contain information about the small-scale nonlinear virial motions.
Finally, the accuracy of the reconstructed velocities affect our forecasts which depend on alignment of the filters based on our knowledge of the true velocities. 
Thanks to our normalization Eqs.~\eqref{vw} and \eqref{eq:estimator_mean} of the moving lens estimator by $1/r_{v_\perp}$,
the expected signal is independent of $r_{v_\perp}$, while the uncertainty on it scales as $1/\left(r_{v_\perp} \sqrt{N_{\textcolor{red}{\text{gal}}}}\right)$. 
As a result, our SNR forecasts in the shot noise regime will scale as 
$ r_{v_\perp} \sqrt{N_{\textcolor{red}{\text{gal}}}}$. 
The better the velocity reconstruction method performs and larger the number of galaxies in the stacking sample we use, the better signal to noise ratio is obtained. 
Refs.~\cite{2023arXiv231212435R,2023arXiv231212434H} show that the correlation coefficient of about 0.8 between true and reconstructed velocities can be expected in an LSST like survey.
They also investigate the effect of photo-z errors on the velocity reconstruction and show that the transverse component of the velocities is less sensitive to the uncertainty in photo-z's, so the moving lens measurements SNR is less penalized by photo-z uncertainties than kSZ.

\section{Conclusion and Outlook}\label{conclusion}

 A galaxy cluster moving with a non zero velocity across the line of sight induces a temperature anisotropy onto the cosmic microwave background. 
 This effect, smaller in size than other secondary anisotropies like tSZ, kSZ or lensing, has a unique signature as a dipole aligned with the direction of the transverse motion.
 While this effect has not yet been detected, our study confirms previous results \cite{2019ApJ...873L..23Y,2019PhRvL.123f1301H,2021PhRvD.103d3536H,2021PhRvD.104h3529H,HotPaol} that it may be marginally detectable with existing datsets, 
and should be measurable at high precision with upcoming experiments such as Simons Observatory/CMB-S4 and DESI/LSST.

In this paper, we generated simulated maps of the moving lens effect
and implemented a moving lens estimator, using stacking weighted by the velocities reconstructed from the galaxy number density field.
We release this as part of the public \texttt{ThumbStack} package which has been used in ACT measurements of tSZ and kSZ effects, and patchy screening forecasts ~\cite{2016PhRvD..93h2002S,2021PhRvD.103f3513S,2024arXiv240113040S}. 

We then performed realistic simulated analyses of the moving lens effect for upcmoing surveys.
We showed that the 2-halo term in the signal is negligible, once high-pass filtering is applied to the maps.
We forecasted the statistical SNR in moving lens for upcoming experiments, finding that the combination of ACT and DESI photometric galaxies data may be sufficient to detect the effect.
Future galaxy surveys with much larger samples will dramatically increase the SNR.

We confirmed the foreground biases in the oriented stacking analysis of the moving lens signal \cite{2023PhRvD.108h3508H}, coming from the large-scale gradients in the tSZ and CIB emissions around the more massive and clustered objects in the CMB map. 
These biases are found to be most significant at higher redshift and for more massive objects.
We also found that the ILCs which null specific foregrounds (e.g., CIB or tSZ) incur an increase in noise power spectrum which dramatically reduces the statistical SNR, making these methods impractical.
We proposed a simple new mitigation method, based on template deprojection, i.e. the removal of the cross correlations with galaxies from the temperature map.
\textcolor{violet}{We showed that this method, in combination with masking and multifrequency cleaning, should reduce foregrounds to an acceptable level for LRG-like galaxies at low redshift. 
We leave a detailed study of foreground mitigation at higher redshift to future work.}

Lastly, we presented a discussion on other noise sources such as the cluster lensing signal, as well as potential systematics such as velocity reconstruction accuracy, photo-z errors and the impact of miscentering of the filters in this measurement.

The moving lens effect is a truly unique probe of velocities across the LOS for extragalactic objects.
Combined with radial velocity information from e.g., kSZ or redshift-space distortions, it will give us access to the full 3D velocity field.
With this 3D velocity information, one can separate the scalar and vector components of the velocities and study late-time vorticity modes~\cite{2023PhRvD.108l3528C}. 

Much like kSZ, the resulting velocity information is a probe of the matter density on the largest scales, and should allow us to constrain the degree of non-Gaussianity in the primordial fluctuations ($f_{\textcolor{red}{\text{NL}}}$) \cite{2018arXiv181013423S,PhysRevD.100.083508}. 

Unlike kSZ, the moving lens is a purely gravitational effect, rather than a scattering effect. 
This distinction is crucial because it means the moving lens effect is not influenced by the complexities of baryonic physics, such as gas density, metallicity and ionization fraction within clusters, which are degenerate with the amplitude of velocities in kSZ measurements. 
Consequently, the moving lens effect may offer a cleaner and more direct measurement of cosmic velocities, allowing us to break the optical depth-velocity degeneracy inherent in kSZ measurements.
This should yield the growth rate of structure and enable tests of gravity \cite{2021PhRvD.104h3529H,2022arXiv220513015H}.
Finally, since the moving lens effect depends on velocities across the LOS, rather than along the LOS for kSZ, it is less affected by photometric redshift errors~\cite{2023arXiv231212435R}, and thus well suited to future large photometric surveys like Rubin LSST, Euclid and Roman.
The moving lens effect thus will enhance our ability to test fundamental cosmological theories with greater precision.
%

\section*{Acknowledgments}

We thank William Coulton for providing noise curves, Abhishek Maniyar for halo model calculations of the tSZ and CIB power spectra, Theo Schutt and Bernardita Ried Guachalla for providing LSST galaxies' host halo mass distributions, and Simone Ferraro, Selim Hotinli, Elena Pierpaoli, and Delon Shen for useful discussions. 
AB thanks Yilun Guan and Emily Biermann for helpful discussions in the early stages of this work. 
The sky simulations used in this paper were developed by the WebSky Extragalactic CMB Mocks team, with the continuous support of the Canadian Institute for Theoretical Astrophysics (CITA), the Canadian Institute for Advanced Research (CIFAR), and the Natural Sciences and Engineering Council of Canada (NSERC), and were generated on the Niagara supercomputer at the SciNet HPC Consortium \cite{2019arXiv190713600P}. 
SciNet is funded by the Canada Foundation for Innovation under the auspices of Compute Canada; the Government of Ontario; Ontario Research Fund - Research Excellence; and the University of Toronto.
%

\bibliographystyle{prsty.bst}
\bibliography{refs}

\begin{thebibliography}{10}

\bibitem{2008RPPh...71f6902A}
N. {Aghanim}, S. {Majumdar}, and J. {Silk}, Reports on Progress in Physics {\bf
  71},  066902  (2008).

\bibitem{2006PhR...429....1L}
A. {Lewis} and A. {Challinor}, \physrep {\bf 429},  1  (2006).

\bibitem{1967ApJ...147...73S}
R.~K. {Sachs} and A.~M. {Wolfe}, \apj {\bf 147},  73  (1967).

\bibitem{1968Natur.217..511R}
M.~J. {Rees} and D.~W. {Sciama}, \nat {\bf 217},  511  (1968).

\bibitem{1972CoASP...4..173S}
R.~A. {Sunyaev} and Y.~B. {Zeldovich}, Comments on Astrophysics and Space
  Physics {\bf 4},  173  (1972).

\bibitem{1980ARA&A..18..537S}
R.~A. {Sunyaev} and I.~B. {Zeldovich}, \araa {\bf 18},  537  (1980).

\bibitem{1980MNRAS.190..413S}
R.~A. {Sunyaev} and Y.~B. {Zeldovich}, \mnras {\bf 190},  413  (1980).

\bibitem{2013ApJ...778...52S}
J. {Sayers} {\it et~al.}, \apj {\bf 778},  52  (2013).

\bibitem{PhysRevLett.109.041101}
N. Hand {\it et~al.}, Phys. Rev. Lett. {\bf 109},  041101  (2012).

\bibitem{Bernardis_2017}
F.~D. Bernardis {\it et~al.}, Journal of Cosmology and Astroparticle Physics
  {\bf 2017},  008  (2017).

\bibitem{PhysRevD.104.043502}
V. Calafut {\it et~al.}, Phys. Rev. D {\bf 104},  043502  (2021).

\bibitem{PhysRevD.103.063513}
E. Schaan {\it et~al.}, Phys. Rev. D {\bf 103},  063513  (2021).

\bibitem{2018arXiv181013423S}
K.~M. {Smith} {\it et~al.}, arXiv e-prints  arXiv:1810.13423  (2018).

\bibitem{1983Natur.302..315B}
M. {Birkinshaw} and S.~F. {Gull}, \nat {\bf 302},  315  (1983).

\bibitem{1986Natur.324..349G}
L.~I. {Gurvits} and I.~G. {Mitrofanov}, \nat {\bf 324},  349  (1986).

\bibitem{1993ApJ...415..459P}
T. {Pyne} and M. {Birkinshaw}, \apj {\bf 415},  459  (1993).

\bibitem{1998A&A...334..409A}
N. {Aghanim}, S. {Prunet}, O. {Forni}, and F.~R. {Bouchet}, \aap {\bf 334},
  409  (1998).

\bibitem{2007ApOpt..46.3444F}
J.~W. {Fowler} {\it et~al.}, \ao {\bf 46},  3444  (2007).

\bibitem{2011ApJS..194...41S}
D.~S. {Swetz} {\it et~al.}, \apjs {\bf 194},  41  (2011).

\bibitem{2016ApJS..227...21T}
R.~J. {Thornton} {\it et~al.}, \apjs {\bf 227},  21  (2016).

\bibitem{2016JLTP..184..772H}
S.~W. {Henderson} {\it et~al.}, Journal of Low Temperature Physics {\bf 184},
  772  (2016).

\bibitem{2019JCAP...02..056A}
P. {Ade} {\it et~al.}, \jcap {\bf 2019},  056  (2019).

\bibitem{2016arXiv161002743A}
K.~N. {Abazajian} {\it et~al.}, arXiv e-prints  arXiv:1610.02743  (2016).

\bibitem{2014ApJS..211...17A}
C.~P. {Ahn} {\it et~al.}, \apjs {\bf 211},  17  (2014).

\bibitem{2017MNRAS.464.1640S}
A.~G. {S{\'a}nchez} {\it et~al.}, \mnras {\bf 464},  1640  (2017).

\bibitem{2017MNRAS.470.2617A}
S. {Alam} {\it et~al.}, \mnras {\bf 470},  2617  (2017).

\bibitem{2016arXiv161100036D}
{DESI Collaboration} {\it et~al.}, arXiv e-prints  arXiv:1611.00036  (2016).

\bibitem{2009arXiv0912.0201L}
{LSST Science Collaboration} {\it et~al.}, arXiv e-prints  arXiv:0912.0201
  (2009).

\bibitem{2018arXiv180901669T}
{The LSST Dark Energy Science Collaboration} {\it et~al.}, arXiv e-prints
  arXiv:1809.01669  (2018).

\bibitem{2022ARA&A..60..363N}
J.~A. {Newman} and D. {Gruen}, \araa {\bf 60},  363  (2022).

\bibitem{2023arXiv231212435R}
B. {Ried Guachalla}, E. {Schaan}, B. {Hadzhiyska}, and S. {Ferraro}, arXiv
  e-prints  arXiv:2312.12435  (2023).

\bibitem{2019ApJ...873L..23Y}
S. {Yasini}, N. {Mirzatuny}, and E. {Pierpaoli}, \apjl {\bf 873},  L23  (2019).

\bibitem{2004A&A...419..439R}
J.~A. {Rubi{\~n}o-Mart{\'\i}n}, C. {Hern{\'a}ndez-Monteagudo}, and T.~A.
  {En{\ss}lin}, \aap {\bf 419},  439  (2004).

\bibitem{2007A&A...467..411M}
M. {Maturi}, T. {En{\ss}lin}, C. {Hern{\'a}ndez-Monteagudo}, and J.~A.
  {Rubi{\~n}o-Mart{\'\i}n}, \aap {\bf 467},  411  (2007).

\bibitem{2019PhRvL.123f1301H}
S.~C. {Hotinli} {\it et~al.}, \prl {\bf 123},  061301  (2019).

\bibitem{2021PhRvD.103d3536H}
S.~C. {Hotinli}, M.~C. {Johnson}, and J. {Meyers}, \prd {\bf 103},  043536
  (2021).

\bibitem{2021PhRvD.104h3529H}
S.~C. {Hotinli}, K.~M. {Smith}, M.~S. {Madhavacheril}, and M. {Kamionkowski},
  \prd {\bf 104},  083529  (2021).

\bibitem{HotPaol}
S. Hotinli and E. Pierpaoli, Journal of Cosmology and Astroparticle Physics
  {\bf 2024},    (2024).

\bibitem{2021PhRvD.103f3513S}
E. {Schaan} {\it et~al.}, \prd {\bf 103},  063513  (2021).

\bibitem{2020JOSS....5.2608Y}
S. {Yasini} {\it et~al.}, The Journal of Open Source Software {\bf 5},  2608
  (2020).

\bibitem{2023PhRvD.108h3508H}
S.~C. {Hotinli}, E. {Pierpaoli}, S. {Ferraro}, and K. {Smith}, \prd {\bf 108},
  083508  (2023).

\bibitem{2019sgai.book.....C}
S.~M. {Carroll}, \textit{Spacetime and Geometry: An Introduction to General
  Relativity}, Addison-Wesley  (2003).

\bibitem{1997ApJ...490..493N}
J.~F. {Navarro}, C.~S. {Frenk}, and S.~D.~M. {White}, \apj {\bf 490},  493
  (1997).

\bibitem{2023arXiv231212434H}
B. {Hadzhiyska}, S. {Ferraro}, B. {Ried Guachalla}, and E. {Schaan}, arXiv
  e-prints  arXiv:2312.12434  (2023).

\bibitem{2017MNRAS.467.2331V}
M. {Vargas-Maga{\~n}a}, S. {Ho}, S. {Fromenteau}, and A.~J. {Cuesta}, \mnras
  {\bf 467},  2331  (2017).

\bibitem{2016PhRvD..93h2002S}
E. {Schaan} {\it et~al.}, \prd {\bf 93},  082002  (2016).

\bibitem{2023JCAP...11..097Z}
R. {Zhou} {\it et~al.}, \jcap {\bf 2023},  097  (2023).

\bibitem{2022PhRvD.106l3517K}
A. {Kusiak}, B. {Bolliet}, A. {Krolewski}, and J.~C. {Hill}, \prd {\bf 106},
  123517  (2022).

\bibitem{dc2}
{LSST Dark Energy Science Collaboration (LSST DESC)} {\it et~al.}, \apjs {\bf
  253},  31  (2021).

\bibitem{2024arXiv240113040S}
T. {Schutt} {\it et~al.}, arXiv e-prints  arXiv:2401.13040  (2024).

\bibitem{SOScience}
{Simons Observatory Collaboration} {\it et~al.}, {\jcap} {\bf 2019},  056
  (2019).

\bibitem{2019MNRAS.483.2236S}
G. {Stein}, M.~A. {Alvarez}, and J.~R. {Bond}, \mnras {\bf 483},  2236  (2019).

\bibitem{2020JCAP...10..012S}
G. {Stein} {\it et~al.}, \jcap {\bf 2020},  012  (2020).

\bibitem{2023PhRvD.108l3501K}
A. {Kusiak}, K.~M. {Surrao}, and J.~C. {Hill}, \prd {\bf 108},  123501  (2023).

\bibitem{2015ApJ...806..247B}
E.~J. {Baxter} {\it et~al.}, \apj {\bf 806},  247  (2015).

\bibitem{PhysRevLett.123.181301}
S. Raghunathan {\it et~al.}, Phys. Rev. Lett. {\bf 123},  181301  (2019).

\bibitem{HuDeDeo}
W. {Hu}, S. {DeDeo}, and C. {Vale}, New Journal of Physics {\bf 9},  441
  (2007).

\bibitem{2012ApJ...757....2G}
M.~R. {George} {\it et~al.}, \apj {\bf 757},  2  (2012).

\bibitem{2023PhRvD.108l3528C}
W.~R. {Coulton}, K. {Akitsu}, and M. {Takada}, \prd {\bf 108},  123528  (2023).

\bibitem{PhysRevD.100.083508}
M. M\"unchmeyer {\it et~al.}, Phys. Rev. D {\bf 100},  083508  (2019).

\bibitem{2022arXiv220513015H}
M. {Howard}, A. {Kosowsky}, and G. {Valogiannis}, arXiv e-prints
  arXiv:2205.13015  (2022).

\bibitem{2019arXiv190713600P}
M. {Ponce} {\it et~al.}, arXiv e-prints  arXiv:1907.13600  (2019).

\end{thebibliography}

\newpage
\appendix

\section{Impact of NFW truncation radius on the radial profile of the moving lens signal}
\label{app:nfw_truncation_radius}

\begin{figure}[h]
    \centering
    \includegraphics[width=.4\textwidth]{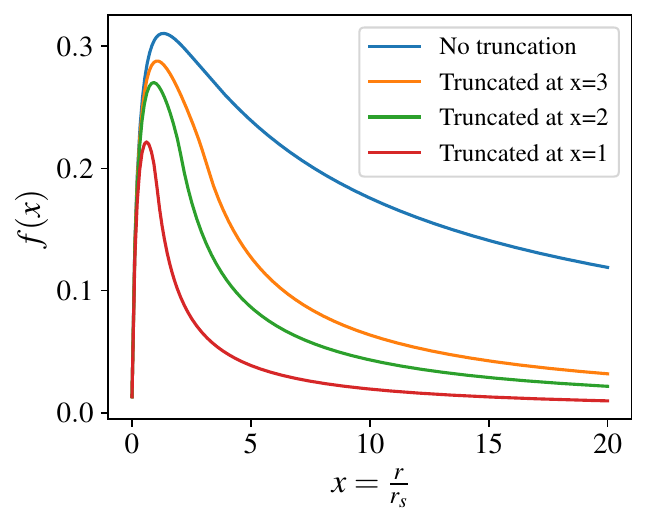}\\
    \caption{Plot of f(x) for the untruncated and truncated NFW mass density.
    }
    \label{fig:f(x)}
\end{figure}

If we truncate the NFW density profile at $x = r/r_s = c_{\rm tr}$, the radial profile $f(x)$ of Eq.~(\ref{signal_final}) for the moving lens signal has the form
\begin{equation}\label{f_trunc}
\begin{split}
    f(x) &=    
         \frac{-x}{(1+c_{\text{tr}})(c_{\text{tr}}+\sqrt{c_{\text{tr}}^2-x^2})}\\ 
         &+ \frac{\ln(1+c_{\text{tr}})+\ln(x)-\ln(c_{\text{tr}}+\sqrt{c_{\text{tr}}^2-r^2})}{x} \\ 
        &-\frac{\ln(-c_{\text{tr}}-x^2+\sqrt{1-x^2}\sqrt{c_{\text{tr}}^2-x^2})}{x\sqrt{1-x^2}} \\ 
         & + \frac{\ln(1+c_{\text{tr}})+\ln(-x)}{x\sqrt{1-x^2}}
\end{split}   
\end{equation}
for $x<c_{\text{tr}}$, and 
\begin{equation}
f(x)
=\frac{-c_{\text{tr}}/(1+c_{\text{tr}})+\ln(1+c_{\text{tr}})}{x}
\end{equation}
for  $c_{\text{tr}}<x$.
Fig.~\ref{fig:f(x)} shows the radial dependence of the moving lense signal $f(x)$ for various truncation radii of the NFW mass density profile.

\section{Derivation of velocity weighted estimator and comparison with uniform weighted estimator}
\label{app:vwuwCompar}

In this section we first provide a derivation of the moving lens estimator of Eq.~\eqref{vw}. Let us begin by writing an estimator for the lensing deflection around any object $j$ as follows
\beq
\hat{\beta}_{j} \coloneqq \frac{\mathcal{T}_j}{v^\text{\textcolor{red}{rec}}_{\perp,j}/c}
\eeq
\textcolor{red}{where $v_\perp$ is the modulus of the 2D perpendicular velocity.}
We want to linearly combine the $\hat{\beta}_{j}$ estimators in a way that is unbiased and has minimum variance. So we write the estimator $\hat{\beta}$ as
\beq
\hat{\beta} = \frac{\sum_j w_j \hat{\beta}_j}{\sum_j w_j}
\eeq
where $w_i = 1/\text{Var}(\beta_j)$ are the inverse variance weightings. If we ignore the map depth variations we can approximate $\text{Var}(\mathcal{T}_j)\approx\sigma^2_T$ for all objects. \textcolor{red}{Furthermore, to compute the variance of our estimator, it is sufficient to consider a fixed realization of the reconstructed velocities, and compute the variance as one varies the realization of the temperature noise map. In this calculation, the reconstructed velocity of each object is thus fixed, and is no longer a random variable. Additionally, as discussed in Sec.~\ref{noise}, in our noise simulations the temperature map has effectively no velocity information in it, thus can they can be assumed independent.} Hence:
\begin{align}
\text{Var}(\hat{\beta}_j) &= \text{Var}\left(\frac{\mathcal{T}_j}{v^\text{\textcolor{red}{rec}}_{\perp,j}}\right) = \frac{\text{Var}(\mathcal{T}_j)}{(v^\text{\textcolor{red}{rec}}_{\perp,j})^2} \nonumber \\
&\approx \frac{\sigma^2_\mathcal{T}}{(v^\text{\textcolor{red}{rec}}_{\perp,j})^2} \propto \frac{1}{(v^\text{\textcolor{red}{rec}}_{\perp,j})^2}.
\end{align}
Therefore we can write
\begin{align}
\hat{\beta} &= \frac{\sum_j w_j \hat{\beta}_j}{\sum_j w_j} 
= \frac{\sum_j \frac{\hat{\beta_j}}{\text{Var}(\hat{\beta}_j)} }
{\sum_j \frac{1}{\text{Var}(\hat{\beta}_j)}} \nonumber \\
&\approx \frac{\sum_j \frac{\mathcal{T}_j}{v^\text{\textcolor{red}{rec}}_{\perp,j}} (v^\text{\textcolor{red}{rec}}_{\perp,j})^2 }
{\sum_j (v^\text{\textcolor{red}{rec}}_{\perp,j})^2}
= \frac{\sum_j \mathcal{T}_j v^\text{\textcolor{red}{rec}}_{\perp,j}}
{\sum_j (v^\text{\textcolor{red}{rec}}_{\perp,j})^2}.
\end{align}
Now if we wish to convert this deflection angle estimator into \textcolor{red}{an} estimator for the moving lens temperature \textcolor{violet}{anisotropy} with units of typical temperature, we can define 
\begin{align} \label{eq:ml_estimator}
\hat{T}_{\text{ML}} &\coloneqq \frac{v_{\perp,\text{rms}}^\text{\textcolor{red}{rec}}}{c} \hat{\beta} \nonumber \\
&= \frac{v_{\perp,\text{rms}}^\text{\textcolor{red}{rec}}}{c} \frac{\sum_j \mathcal{T}_j v_{\perp,j}}
{\sum_j v_{\perp,j}^2}
\end{align}
which is what we have in Eq.~\eqref{vw} assuming $r_{v_{\perp}}=1$. However, in practice $r_{v_{\perp}}< 1$. In order to make the expectation value of estimator Eq.~\eqref{eq:ml_estimator} independent of $r_{v_{\perp}}$ we normalize it by the prefactor $1/r_{v_{\perp}}$ and obtain Eq.~\eqref{vw}. 
Now we wish to compare the signal to noise ratio obtained from the velocity-weighted estimator of Eq.~(\ref{vw}) with a uniformly-weighted estimator for the moving lens signal such as
\beq \label{uw}
\hat{T}_{\text{ML}} =  \frac{\sum_j^N \mathcal{T}_j}{N}.
\eeq
For simplicity we assume clusters have the same masses and redshifts, so the lensing deflection angle will be the same for all, $\bm{\beta}_j \equiv \bm{\beta}$. In the case of a uniformly-weighted estimator of Eq.~(\ref{uw}), the estimated signal $S_{U}$ will be of the form
\beq
S_{U} 
=
\beta \frac{1}{N} \sum_j^N
v_{\textcolor{red}{\perp},j}^\text{\textcolor{red}{rec}}
=
\beta \langle v_{\textcolor{red}{\perp}}^\text{\textcolor{red}{rec}} \rangle
\eeq
and the variance of the signal will be
\beq
N_{U}^2 
=
\frac{1}{N^2} \sum_j^N \sigma_\mathcal{T}^2 = \frac{\sigma_\mathcal{T}^2}{N}.
\eeq
Thus the signal to noise ratio in this case will be 
\beq
\text{SNR}_U
=
\frac{\beta \langle v_{\textcolor{red}{\perp}}^\text{\textcolor{red}{rec}} \rangle \textcolor{violet}{\sqrt{N}}}{\sigma_\mathcal{T} }.
\eeq
On the other hand, for a velocity-weighted estimator as in Eq.~(\ref{vw}) we can write the estimated signal $S_{V}$ as
\beq
S_{V} 
=
\textcolor{violet}{\frac{v_{{\perp,\text{rms}}}^\text{rec}}{c}}
\frac{\sum_j^N \beta v_{\textcolor{red}{\perp},j}^\text{\textcolor{red}{rec}}(v_{{\textcolor{red}{\perp},j}}^\text{\textcolor{red}{rec}})}{\sum_j^N (v_{{\textcolor{red}{\perp},j}}^\text{\textcolor{red}{rec}})^2} = \textcolor{violet}{\frac{v_{{\perp,\text{rms}}}^\text{rec}}{c}}\beta
\eeq
with the variance
\begin{align}
N_{V}^2 
&=
\textcolor{violet}{\left(\frac{v_{{\perp,\text{rms}}}^\text{rec}}{c}\right)^2}
\frac{\sum_j^N \sigma_\mathcal{T}^2 (v_{\textcolor{red}{\perp},j}^\text{\textcolor{red}{rec}})^2}{(\sum_j^N (v_{\textcolor{red}{\perp},j}^\text{\textcolor{red}{rec}})^2)^2} \nonumber \\
&= 
\textcolor{violet}{\left(\frac{v_{{\perp,\text{rms}}}^\text{rec}}{c}\right)^2}
\frac{\sigma_\mathcal{T}^2}{\sum_j^N (v_{\textcolor{red}{\perp},j}^\text{\textcolor{red}{rec}})^2} \nonumber \\
&= 
\textcolor{violet}{\left(\frac{v_{{\perp,\text{rms}}}^\text{rec}}{c}\right)^2}
\frac{\sigma_\mathcal{T}^2}{N (v_{\textcolor{red}{\perp},\text{rms}}^\text{\textcolor{red}{rec}})^2}
\end{align}
Therefore, the signal to noise ratio in this case will be
\beq
\text{SNR}_V 
=
\frac{\beta v_{{\textcolor{red}{\perp},\text{rms}}}^\text{\textcolor{red}{rec}}\textcolor{violet}{\sqrt{N}}}{\sigma_{\mathcal{T}}}
\eeq
Now, if we take the ratio of the signal to noise ratios we have:
 
\textcolor{violet}{{\beq
\frac{\text{SNR}_V}{\text{SNR}_U} \approx \frac{v_{{\perp,\text{rms}}}^\text{rec}}{ \langle v_{\perp}^\text{rec} \rangle} = \frac{2}{\sqrt{\pi}}
\simeq 1.13
\eeq
where we have assumed the two-dimensional $v_\perp^{\text{rec}}$ components are independent gaussian random variables with standard deviation $\sigma$ such that $v_\perp^{\text{rec}}$ follows the 2D Rayleigh distribution, leading to $v_{{\perp,\text{rms}}}^\text{rec}=\sqrt{2}\sigma$ and $\langle v_{\perp}^\text{rec} \rangle = \sqrt{\pi/2}\sigma$.}
So we can conclude that the velocity-weighted estimator indeed improves the signal to noise ratio of the moving lens measurement by 13\%.}

\section{Pipeline validation} \label{validation}

Fig.~\ref{fig:cosrings, map of 2 similar objects} shows the effect of different high-pass filters on the profile of the moving lens signal measured from a single galaxy cluster.
As mentioned in section \ref{noise} we chose to use the $l_{\textcolor{red}{\text{cut}}}=2000$ as it can help with decorrelation of the apertures without causing too much ringing in the profile.
\begin{figure}[h]
    \centering
    \includegraphics[width=.38\textwidth]{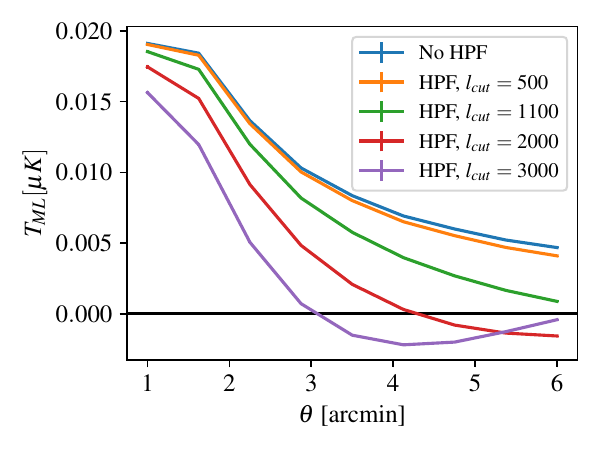}\\
    \caption{The measured moving lens profile of a single galaxy as a function of cosine ring aperture sizes after applying different high-pass filters to the simulated map. 
    }
    \label{fig:cosrings, map of 2 similar objects}
\end{figure}
In order to obtain the stacked profile of the moving lens signal and isolate its 1-halo and 2-halo contributions, we simulated 45$\times$45 deg$^2$ maps of the signal for galaxy clusters with true and shuffled velocities based on the BOSS catalog. To speed up the simulation, we downsampled the catalog by $5\%$ which reduces the 2-halo contribution to the signal but we account for that at the end \textcolor{red}{by multiplying the 2-halo contribution by a factor $20 = 1/0.05$. Indeed, the 2-halo term should scale as the number of neighboring halos around each galaxy, i.e. linearly with the total number of objects.} Also, we chose to generate maps with $0.43$' arcmin resolution which causes discrepancies with the theoretical prediction for the stacked profile in the smaller apertures. The resulting profiles are presented in top panel of Fig.~\ref{fig:rawCurves} below.
\begin{figure}[h]
    \centering
    \includegraphics[width=.47\textwidth]{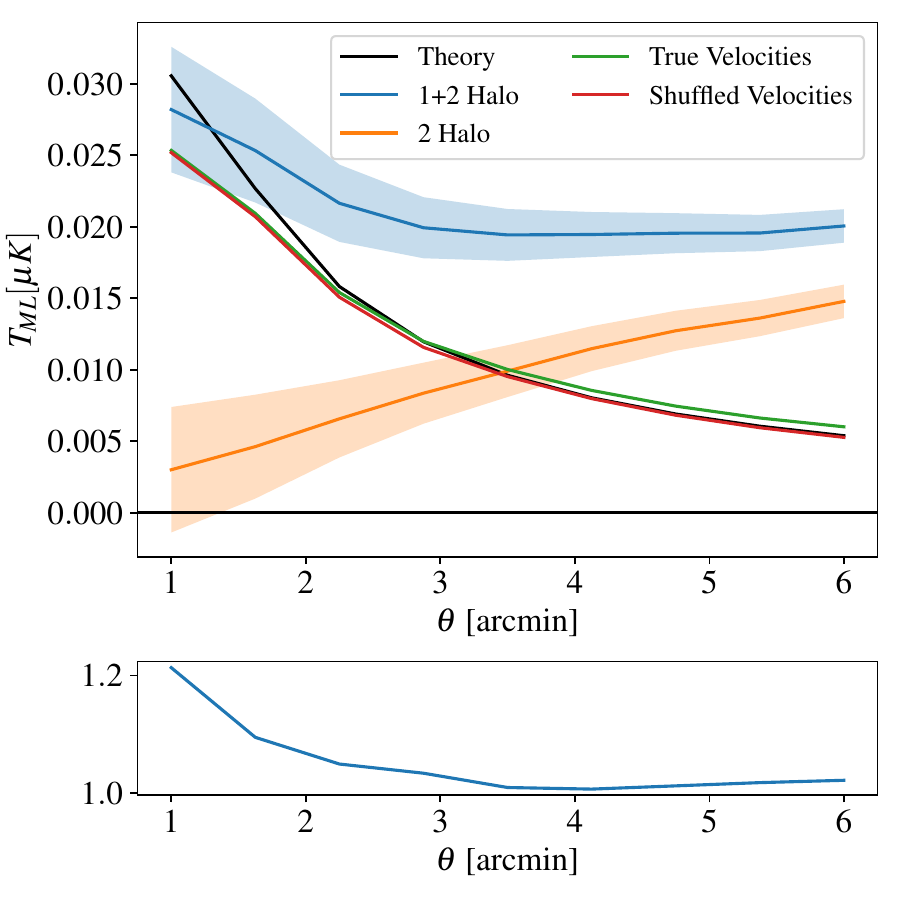}
    \caption{\textbf{Top.} The stacked profiles and comparison with theory prediction for a 5\% down-sampled 45$\times$45 deg$^2$ map of 7977 galaxy clusters with true and shuffled velocities. \textbf{Bottom.} Profile of the transfer function used to correct for the imperfect resolution in the simulated maps. }
    \label{fig:rawCurves}
\end{figure}
We confirmed that the discrepancy between the theoretical prediction and the stacked profile of the shuffled velocities map reduces by simulating maps with better resolution which demand longer time to generate. Thus, we decided to correct for this imperfect pixellation using a transfer function that is equal to the ratio of theoretical stacked profile over the measured profile from the shuffled velocities map. This transfer function is shown in bottom panel of Fig.~\ref{fig:rawCurves}. 
The resulting 1-halo and 2-halo contributions after performing the process above are shown in Fig.~\ref{fig:1-2-halo}. It is worth noting that in the simulations of the 2-halo term contributions, it is crucial to choose a patch of sky large enough such that it samples the velocity field evenly and there is no significant mean transverse velocity in it, otherwise the shuffling of the velocities will not completely null the 2-halo term and it would lead to a component of the 2-halo term in the stacked profile that is originating from the mean velocity of the objects in the patch.
\begin{figure}[h]
    \centering
    \includegraphics[width=.43\textwidth]{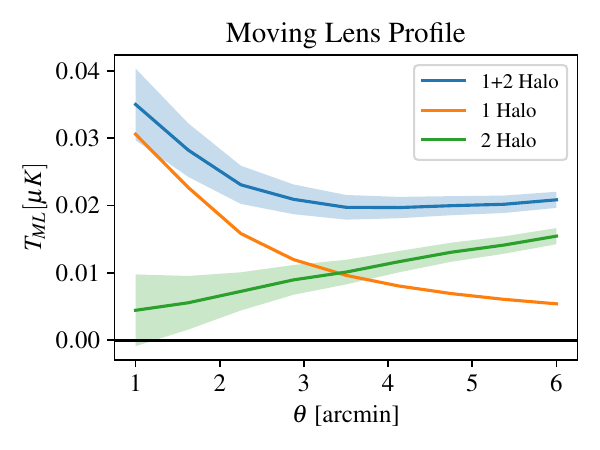}\\
    \caption{The stacked profiles of the 1-halo and 2-halo contributions to the moving lens signal for a simulated 45$\times$45 deg$^2$ patch of galaxy clusters based on BOSS catalog. However, Sec~\ref{aper} shows that the 2-halo term has negligible effect in practice, due to the high pass filter that we need to apply to remove large scale noise of primary anisotropies of CMB.
    }
    \label{fig:1-2-halo}
\end{figure}
Furthermore, we investigated the convergence \textcolor{red}{of} 2-halo term contribution to the signal as a function of the size of the patch of the clusters on which we perform the stacking analysis. It is crucial to make sure that the stacking patch is large enough such that it samples the correlations in the transverse velocity field correctly and also has enough galaxies outside of it in the simulated map such that the measured signals at galaxies inside the patch contains 2-halo contributions from galaxies around it in all directions and we avoid edge effects in our analysis. Fig.~\ref{fig:2h_conv} shows the profile of the 2-halo term contribution to the moving lens signal as a function of the stacking patch. This plot suggests that within the size of the error bars, the convergence is attained for the patches larger than 35$\times$35 deg$^2$. 
\begin{figure}[h]
    \centering
    \includegraphics[width=.38\textwidth]{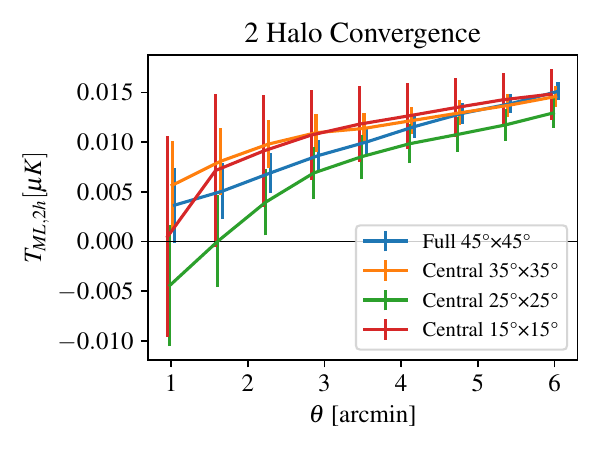}\\
    \caption{Convergence of the 2-halo term contribution to the moving lens signal as a function of the patch size on which the stacking analysis is performed.}
    \label{fig:2h_conv}
\end{figure}
\FloatBarrier

\end{document}